\documentclass[lineno]{jfm}

\usepackage{bm,amsmath,amssymb,mathrsfs,mathtools}
\usepackage{newtxtext}
\usepackage{newtxmath}
\usepackage{xcolor}
\usepackage{graphicx}
\usepackage{subfigure}
\usepackage{bbding}
\usepackage[version=4]{mhchem}
\usepackage[colorlinks=true,urlcolor=blue,linkcolor=blue,citecolor=black,filecolor=blue]{hyperref}
\usepackage{comment}
\usepackage[justification=raggedright,singlelinecheck=false]{caption}

\usepackage{siunitx}

\definecolor{navyblue}{rgb}{0.0, 0.0, 0.5}
\definecolor{olive}{rgb}{0.5, 0.5, 0.0}

\usepackage{multirow}

\newsavebox{\astrutbox}
\sbox{\astrutbox}{\rule[-5pt]{0pt}{20pt}}

\newcommand\nc{\newcommand}
\nc{\vect}[1]{\mbox{\boldmath $#1$}}
\nc{\pt}{\partial_t}
\nc{\px}{\partial_x}

\usepackage{natbib}

\newcommand{\BibitemShut}[1]{}

\graphicspath{{figs/}}

\title{Modeling the dynamics of an oil drop driven by a surface acoustic wave in the underlying substrate}

\shorttitle{Dynamics of a SAW driven drop}

\author{M. Fasano$^1$\thanks{First and second authors did a comparable amount of work, with Fasano focusing on modeling and simulations, and Li on experiments.}, Y. Li$^2$\footnotemark[1],  J.A. Diez$^3$,  J. D'Addesa$^1$, O. Manor$^2$, L.J. Cummings$^1$ and L. Kondic$^1$}

\shortauthor{M. Fasano$^1$\footnotemark[1] et al.}

\affiliation
{
$^1$ Department of Mathematical Sciences and Center for Applied Mathematics and Statistics, New Jersey Institute of Technology, Newark, New Jersey 07102, USA\\
$^2$  Department of Chemical Engineering, Technion - Israel Institute of Technology, Haifa 32000, Israel\\
$^3$  Instituto de F\'{\i}sica Arroyo Seco, Universidad Nacional del Centro de la Provincia de Buenos Aires and CIFICEN-CONICET-CICPBA, 7000 Tandil, Argentina\\
}

\begin{document}
\maketitle

\begin{abstract}
We present a theoretical study, supported by simulations and experiments, on the spreading of a silicone oil drop under MHz-frequency surface acoustic wave (SAW) excitation in the underlying solid substrate. Our time-dependent theoretical model uses the long wave approach and considers interactions between fluid dynamics and acoustic driving. While similar methods have analyzed micron-scale oil and water film dynamics under SAW excitation, acoustic forcing was linked to boundary layer flow, specifically Schlichting and Rayleigh streaming, and acoustic radiation pressure. For the macroscopic drops in this study, acoustic forcing arises from Reynolds stress variations in the liquid due to changes in the intensity of the acoustic field leaking from the SAW beneath the drop and the viscous dissipation of the leaked wave. Contributions from Schlichting and Rayleigh streaming are negligible in this case. Both experiments and simulations show that after an initial phase where the oil drop deforms to accommodate acoustic stress, it accelerates, achieving nearly constant speed over time, leaving a thin wetting layer. Our model indicates that the steady speed of the drop results from the quasi-steady shape of its body. The drop speed depends on drop size and SAW intensity. Its steady shape and speed are further clarified by a simplified traveling wave-type model that highlights various physical effects. Although the agreement between experiment and theory on drop speed is qualitative, the results' trend regarding SAW amplitude variations suggests that the model realistically incorporates the primary physical effects driving drop dynamics. 
\end{abstract}

\section{Introduction}

Dynamic wetting of a solid substrate by a liquid film is a common occurrence. In natural systems, it appears in surfactant films that wet the lungs to promote breathing and the eyes to keep them moist~\citep{veldhuizen_role_2000,holly_wettability_1971}, or in the spreading of water drops on solid surfaces, such as raindrops on windows~\citep{dussan_v_motion_1974,dussan_v_ability_1983}. In technological systems it is harnessed for actuating microfluidic platforms \citep{Atencia:2005bu,Stone:2004kg,Whitesides:2006jj}, for cooling electronic circuits \citep{Amon:2001ji,BarCohen:2006if}, for desalination \citep{Fletcher:1974jz} and for a variety of manufacturing applications~\citep{Fendler:1996gb,Nagayama:1996vd,Wang:2004kx}. Many different mechanisms govern dynamic wetting, where capillary forces at the free surface of a liquid film, or at a meniscus, contribute appreciably to the dynamics of the film. In addition to capillary forces, dynamic wetting may be powered by gravity, thermal or solutal Marangoni effects, or electrokinetics, among other driving mechanisms; see~\citet{oron_rmp97} or \citet{cm_rmp09} for comprehensive reviews. In the present work, we focus on the contributions of mechanical vibrations in the solid substrate, particularly MHz-frequency Rayleigh surface acoustic waves (SAWs), to dynamic wetting~\citep{morozov_vibration-driven_2018}. We study particularly the effects of bulk acoustic streaming -- mass transport in the bulk fluid that results from variations in acoustic intensity therein. While extensively observed and described in the scientific literature \citep{Brunet07,Brunet09,yeoannurev}, we are not aware of a model describing the dynamics of such a system that further accounts for the capillary and gravitational stresses that shape the liquid/air interface for the case when drops are sufficiently thick (millimeter scale, therefore much thicker than the SAW wavelength). Here, we use thin film (long wave) theory alongside physical experiments to develop a better understanding of the dynamics of this problem.   

There are several different mechanisms by which acoustic waves or other mechanical waves, at a solid boundary or at the free surface of a liquid, generate streaming -- steady flow along the wave's path. A mechanical wave at a boundary invokes periodic viscous flow within the viscous penetration length $\delta\equiv\sqrt{2\mu/(\rho\omega)}$ away from the interface, where $\mu,~\rho$ and $\omega$ are the viscosity and density of the liquid, and the wave angular frequency, respectively. Examples in which this problem was considered include the work by Rayleigh on a standing acoustic wave grazing a solid bed~\citep{LordRayleigh1884}; by~\citet{Schlichting:1932p447} on standing vibration of waves of infinite and finite (respectively) wavelength in a solid along its surface; by~\citet{longuet-higgins_mass_1953} on shallow ocean waves; and by~\citet{manor_yeo_friend_2012} on propagating Rayleigh (surface acoustic) waves in a solid substrate. Convective contributions due to surface waves invoke a drift of liquid mass,
which does not attenuate away from the solid surface. This is known as Schlichting streaming~\citep{yeoannurev}, and its component far from the boundary was further coined the ``Rayleigh law of streaming" by \citet{LIGHTHILL:1978p12} in recognition of Rayleigh's work on the drift resulting from the presence of standing acoustic waves in fluid near a solid boundary. 

A separate mechanism, the Eckart streaming, appears regardless of the presence of a boundary. This is an acoustic streaming that occurs in the bulk fluid, and results from {\it variations} in the intensity of sound or ultrasound waves in the fluid. In the classic work by Eckart and in many subsequent studies, acoustic streaming is ascribed to the viscous attenuation of traveling waves in the fluid \citep{Eckart:1948to,Nyborg:2004p312,LIGHTHILL:1978p12}.  This 
attenuation results in spatial variations of the wave intensity in the fluid, and in corresponding spatial variations in convective Reynolds stresses.  The consequence is a secondary flow, whose steady component at long times is the acoustic streaming, characterized by an intense vortical flow field that appears in the bulk fluid. This is the main acoustic mechanism for the actuation of fluid in micro-channels \citep{Wixforth:2003jp,Wixforth:2004p910,yeoannurev,fasano2025phase} and in drop microfluidics \citep{Guttenberg:2004wg,Brunet07,Brunet09,Brunet10}. It usually dominates contributions to the flow in the bulk fluid arising from the boundary layer type acoustic streaming mechanisms, such as Shlichting or Rayleigh streaming described above. An additional effect related to the interaction of acoustic waves with a surface, the acoustic radiation pressure~\citep{hertz_1939,hamilton1998nonlinear}, may be also relevant in the presence of interfaces, such as the fluid-air surface. This effect has been discussed extensively in the literature~\citep{King:1934tp,Campbell:1970wb,chu,Shiokawa:1989tg,Hasegawa:2000vy,Anonymous:5ILrl7tg,PhysRevLett.117.114504,Subramani2023}, and shown capable of deforming and displacing soft interfaces and of governing the dynamics of films \citep{Biwersi:2000ti,Alzuaga05,issenmann_bistability_2006, rajendran_theory_2022, marcos2025Monte}. \citet{altshuler2015spreading} consider the problem of a partially wetting water drop atop a MHz frequency propagating vibration in the substrate, in the setup similar to the one considered here, although the reduced viscosity and increased surface tension lead to reduced viscous dissipation of the leaky SAW in the fluid, and a necessity to include disjoining pressure to model liquid-solid interactions. \citet{Horesh} further consider a vertical setup where gravity becomes important and show that acoustic radiation pressure is negligible for the partially wetting water due to the curvature of the meniscus, which does not support acoustical resonance.

While the literature discussing the interaction of acoustic waves with fluids is clearly extensive, in what follows, we discuss briefly just a few works that are particularly relevant in the present context, in which a SAW traveling in a solid substrate drives flow in a neighboring fluid. In 1970 \citet{campbell_propagation_1970} simulated a SAW in a piezoelectric solid in contact with an ideal (inviscid) fluid. Then, in the late 1980s, \citet{Shiokawa:1989tg} used the ansatz of a harmonic acoustic waveform in the solid based on ideas developed by \citet{Nyborg:2004p312}, converted the approach adopted by~\citet{campbell_propagation_1970} to analytic expressions, and calculated the resulting acoustic forcing in the adjacent inviscid fluid due to attenuation of SAW in the solid. More recently, \citet{vanneste_streaming_2011} revisited the work of~\citet{campbell_propagation_1970} in the context of viscous fluids and calculated numerically the corresponding acoustic forcing in the fluid. 

In the present paper, we focus on developing a model that can be used to formulate a long wave theory applicable to the free surface evolution of thin viscous films and drops due to the presence of a SAW propagating in an underlying solid substrate. To facilitate such a development, we simplify the previously considered approaches in order to isolate the dominant physical effects governing the dynamics of the film.  In particular, we simplify the approach used by~\citet{campbell_propagation_1970} by assuming damped harmonic waves in the solid. Furthermore, we compare the previous approaches by~\citet{vanneste_streaming_2011} and~\citet{Shiokawa:1989tg} to quantify the importance of viscous effects, which were included by the former authors but ignored by the latter. We follow earlier attempts \citep{Brunet10} and focus on the contribution of acoustic streaming to the drop dynamics. In principle, one could also consider including the effects of the acoustic radiation pressure at the free surface, as was done in the recent work of~\citet{marcos2025Monte}, who used a Monte-Carlo type approach to study dynamics of oil-water films under acoustic forcing. However, in the present case, the main contribution to drop displacement is the acoustic streaming, which efficiently transfers power from the SAW to the translational motion of the drop; hence, we do not include acoustic radiation pressure here. This point will be discussed in further detail in Sec.~\ref{sec:model}.

The rest of this paper is structured as follows.  As a motivation, in Sec.~\ref{sec:experiment} we discuss simple physical experiments that demonstrate the well-known phenomenon of thick oil films and drops displacing under the action of MHz-frequency Rayleigh-type SAWs. These experiments illustrate the process of dynamic wetting and drop displacement powered by bulk acoustic streaming. We then derive the theoretical model that can explain such observations in Sec.~\ref{sec:model}. Our theoretical results are presented in Sec.~\ref{sec:results}: we discuss computational results in Sec.~\ref{sec:basic_case}, before analyzing a simplified traveling--wave type model in Sec.~\ref{sec:travel}. Sections~\ref{sec:model} and~\ref{sec:results}, together with the appendices, constitute the main novel contributions of our work.  Section~\ref{sec:conclusion} is devoted to the summary and conclusions. The appendices include 
(\ref{app:RealVel}) details of the theoretical results, (\ref{app:LeakySAW}) various simplifications and approximations including the inviscid limit, and (\ref{app:comsol}) the outline of the numerical approach implemented. Supplementary materials provide videos of selected experiments and animations of representative computational results.

\section{Experiment}
\label{sec:experiment}

The experiments described here are illustrative and portray the well-known phenomenon in which MHz-frequency surface acoustic waves (SAWs) are used to displace drops and thick films of liquid along the path of the SAW. The purpose of these experiments is to demonstrate the system under investigation, highlight key features and trends, and provide results that will be used to motivate the theory and simulations presented in the following sections.

\begin{figure}
\centering
\includegraphics[width=0.40\linewidth]{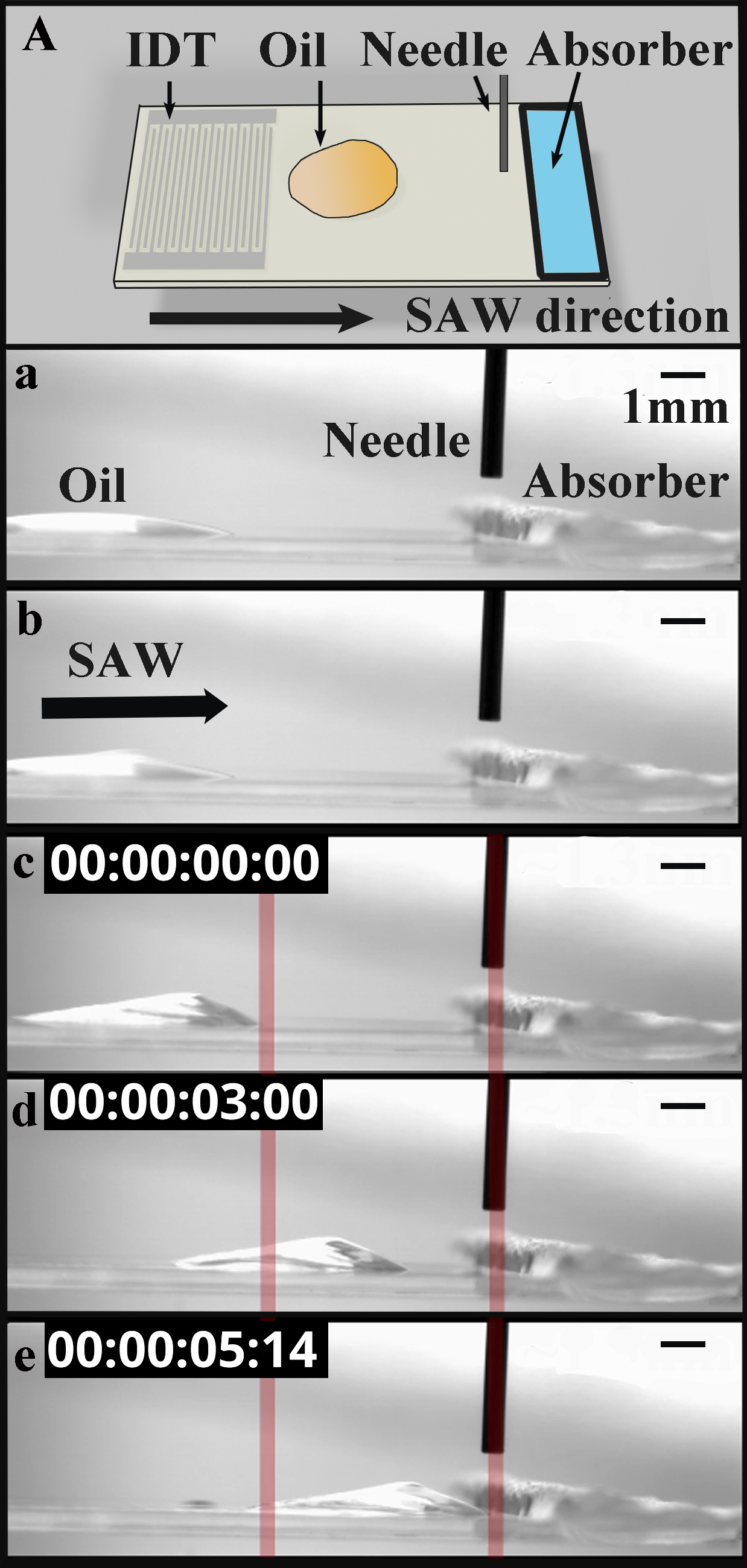}
\caption{(A) Upper schematic view of the experimental setup: Surface acoustic wave (SAW) propagates from the interdigitated transducer (IDT) until it reaches the acoustic absorber (comprised of glycerol-soaked paper placed on the actuator under and to the right of the needle). The needle is of known diameter (510 $\mu $m), placed for identifying spatial resolution in the images.  (a-e) Successive snapshots are taken from an experiment monitoring the flow of a silicone oil film. During the experiment, a drop of silicone oil is placed on the horizontal surface (a), it deforms due to the application of SAW (b), and moves in the direction of SAW propagation (c-e); time is shown in seconds, and the vertical lines serve as a reference.}
\label{fig:exp_saw_intro}
\end{figure}

Figure~\ref{fig:exp_saw_intro} shows the experimental setup (A) and a typical experimental result (a-e).  We generate a propagating 20 MHz frequency SAW by applying a same-frequency sinusoidal voltage signal to a piezoelectric actuator -- a SAW device. The actuator comprises a 5 nm titanium/1 $\mu$m aluminum interdigitated transducer (IDT; from which the SAW emanates) fabricated atop lithium niobate (LiNbO$_3$, Roditi International, UK) by standard lift-off photolithography. The substrate used for the SAW device is 11~mm~$\times$~24~mm in size, 0.5~mm thick, 128° Y-cut, X-propagating, single-crystal piezoelectric lithium niobate, where X and Y are crystal axes  \citep{campbell1968method}. The actuator is integrated into the external electrical signal using pogo pins (BC201403AD, Interconnect Devices, Inc.) assembled in a 3D-printed elastomeric stage, which holds the actuator and is connected to a signal generator (R\&S SMB100A microwave signal generator) and amplifier (model A10160, Tabor Electronics Ltd.).  Prior to the experiment we clean the SAW device using four different solvents: acetone (AR-b, 99.8\%, 67-641, Bio-Lab Ltd.), 2-propanol (AR-b, 99.8\%, 67-63-0, Bio-Lab Ltd.), ethanol (CP-p, 96\%, 64-17-5, Bio-Lab Ltd.), and water (HPLC plus, 7732-18-5, Sigma-Aldrich), before drying the actuator using compressed air. We place a paper cylinder soaked with glycerol at the far end of the actuator from the IDT to absorb the SAW and prevent reflections.

In our experiments, we represent the SAW strength (intensity) by measuring the normal displacement amplitude, $A_{\rm n}$, at the solid surface before any attenuation in the liquid, so that $A_{\rm n}$ is the amplitude of the transverse component of the SAW in the substrate. We measure $A_{\rm n}$ over a surface of $1\times1~\text{mm}^2$ about $2$~mm away from the SAW actuator using a scanning laser Doppler vibrometer (MSA-500, Polytech). We control this amplitude by changing the applied voltage, $V$, of the signal generator at the surface of the actuator:   
consistent with the literature~\citep{ballantine1996acoustic}, there is a linear relationship between $A_{\rm n}$ and $V$; see~\cite{arxiv} for more details. 

\begin{figure}
\centering
\includegraphics[width=0.85\linewidth]{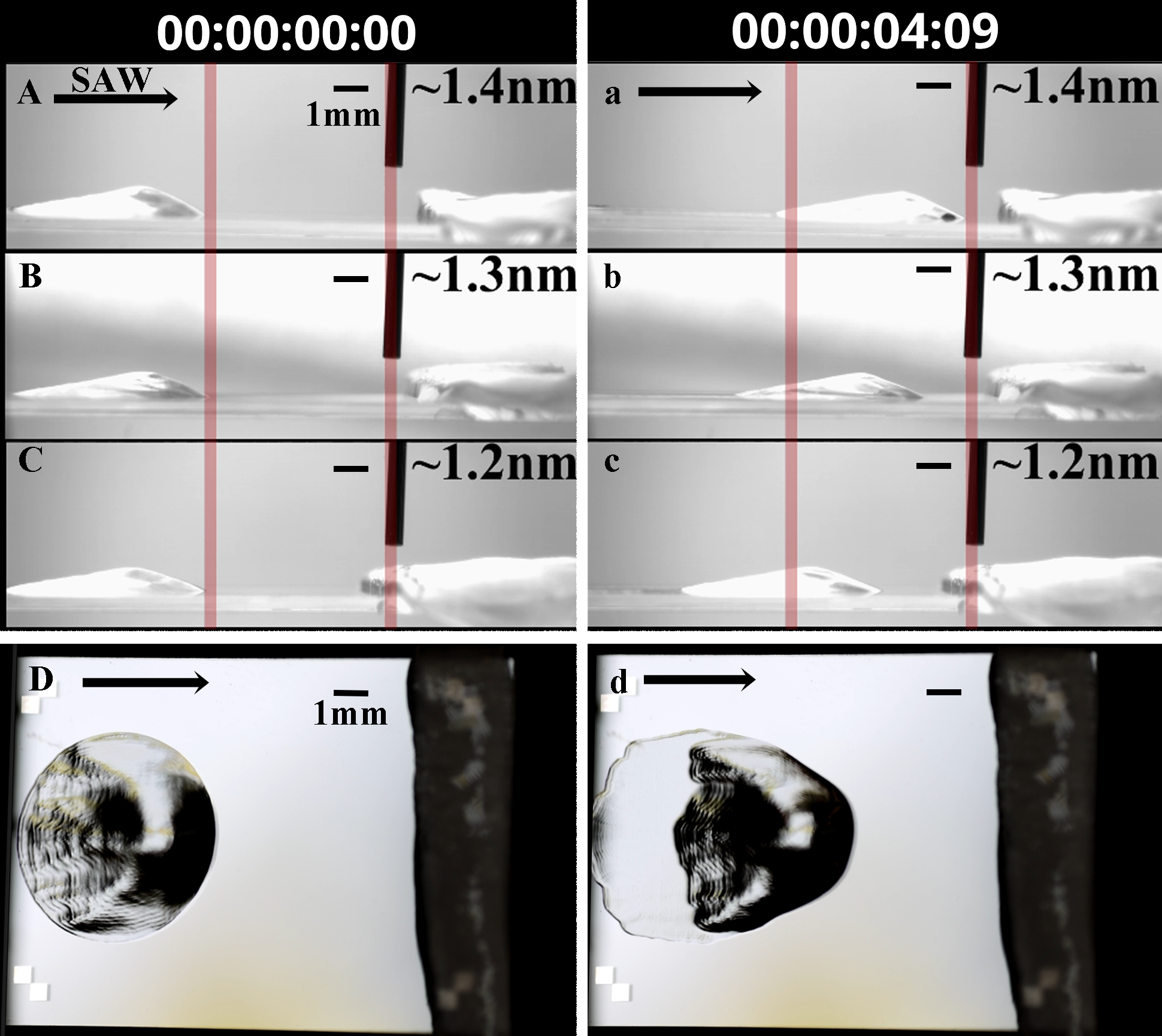}
\caption{Two sets of images from experiments monitoring silicone oil drops powered by SAWs of three different measured amplitudes $A_{\rm n}$ (1.2 nm--1.4 nm), where (A-C) silicone oil starts from the same location, moves in the direction of the SAW, and reaches different distances (a-c) after 4.09 seconds. The corresponding side-view videos at different acoustic power levels are used to analyze the speed and profile of silicone oil during the movement. (D, d) show top view snapshots (corresponding to (C, c)) of the oil.  A thin film of oil, hardly visible in the side view, can be seen more clearly behind the main body of silicone oil in the top views.}
\label{fig:exp_saw}
\end{figure}

 We place an $8~\mu$l (8 mm$^3$) drop of silicone oil (50 cSt, 378356, Sigma-Aldrich) atop the actuator using a pipette, approximately 3.5~mm away from the IDT, and introduce electrical signal at different voltage levels to induce motion of the oil film, which dynamically wets the solid substrate along the path of the SAW. Figure~\ref{fig:exp_saw}(a-c) shows a time-lapse (side view) of the typical silicone oil dynamics, captured using the framework and camera of a goniometer (Data Physics; OCA 15Pro). Figure~\ref{fig:exp_saw}(d) shows a top view obtained using a camera (EOS R5, Canon) with a macro lens (RF 100~mm F2.8L MACRO IS USM, Canon). 
\begin{figure}
\centering
\subfigure[]{
\includegraphics[width=0.485\linewidth]{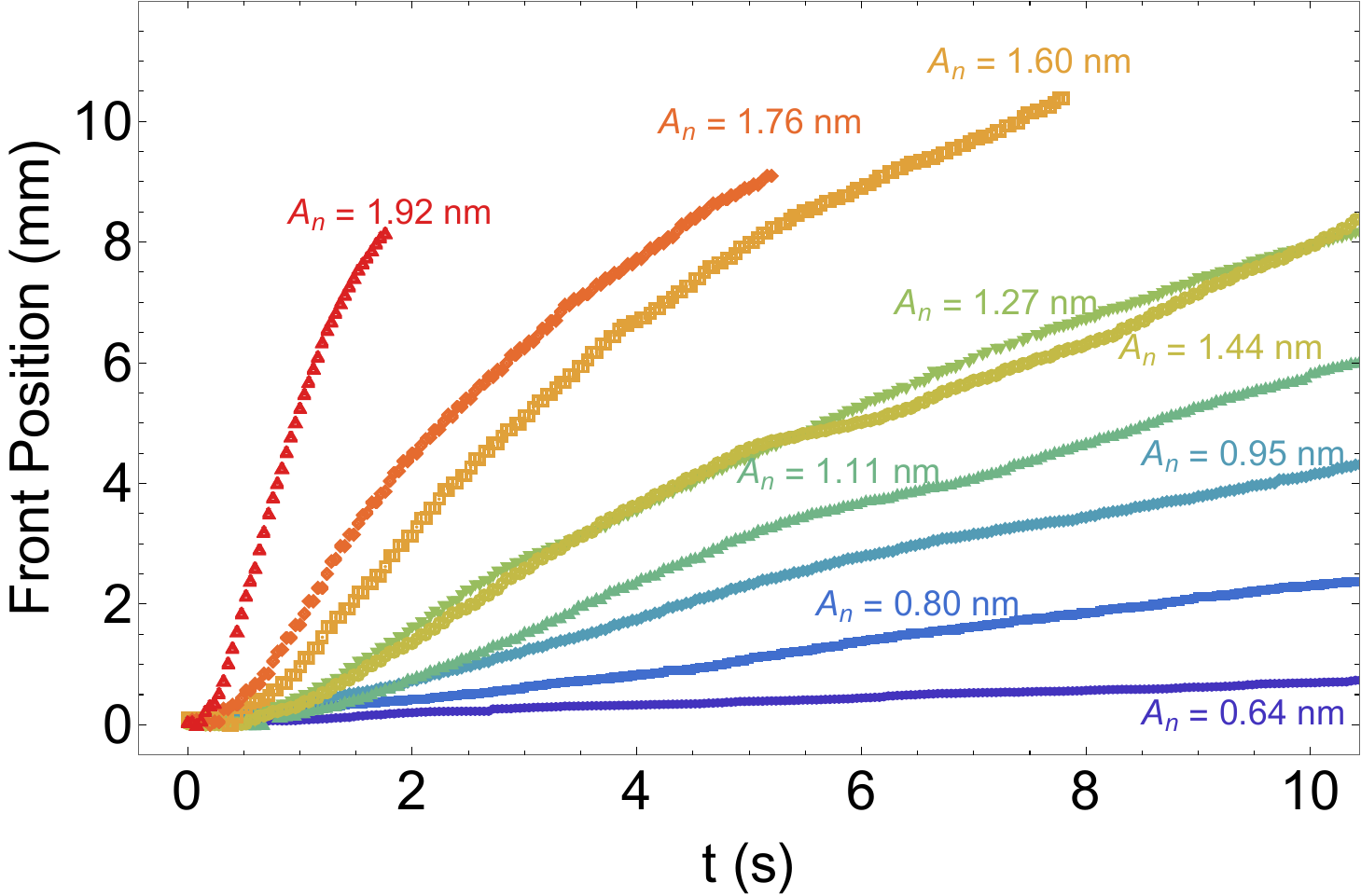}}
\subfigure[]{
\includegraphics[width=0.485\linewidth]{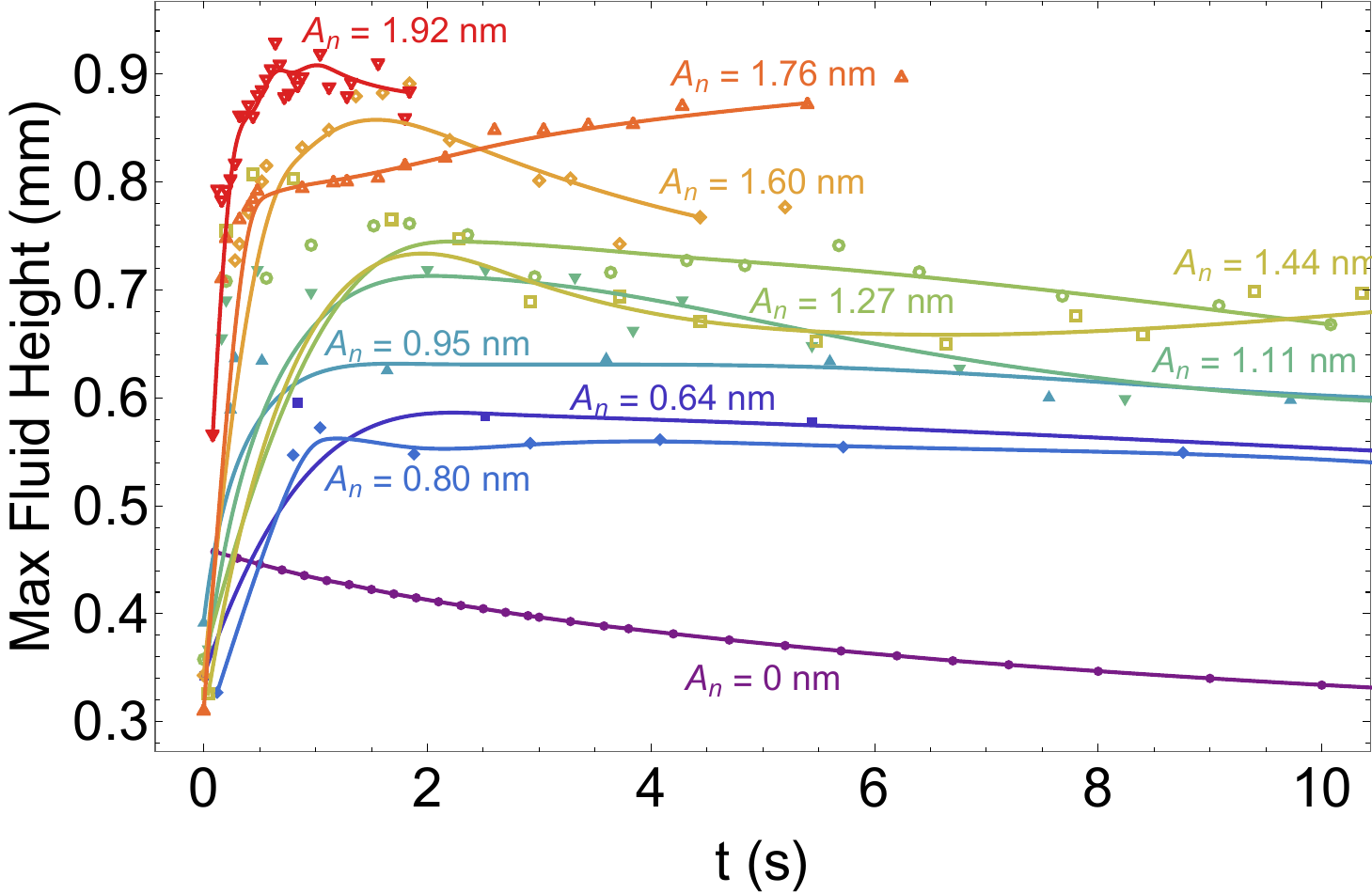}}
\caption{Time evolution of (a) the front position and (b) the maximum drop height for several values of $A_{\rm n}$, for silicone oil drops of volume $\mathcal{V}_{\rm d}=8\, \mu$l and kinematic viscosity $\nu=50$ cSt. The symbols correspond to the measurements, and the lines are simple fits to guide the eye. We expect errors of $\pm 50~\mu$m in the values of the film height due to the limited resolution of the side view image and tracker software. Note that $t=0$ corresponds to the time instant at which SAW is applied, which is 2 s after deposition of the oil. 
A video showing the evolution at various SAW amplitudes is available as a supplementary material (movie 1).}
\label{fig:exp_saw_results2}
\end{figure}
To confirm that the observed dynamics are indeed induced by SAW, we also carried out control experiments, 
in which SAW is absent and all other parameters are kept the same. The oil drop in this experiment, which was repeated three times, slowly spreads concentrically, as expected in the absence of a directional forcing mechanism (figure not shown for brevity). 

In the experiments with SAW actuation we employ the side view video, recorded by a goniometer camera, to capture the position of the advancing three-phase contact line of the SAW-actuated oil film using the open-source software \emph{Tracker} \citep{brown2009innovative} (see Fig.~\ref{fig:exp_saw_results2}(a) for typical data). We also measure the maximum height of the moving oil film using \emph{Tracker} (typical results in Fig.~\ref{fig:exp_saw_results2}(b)).  Figure~\ref{fig:exp_saw_results2}(a) shows that, as expected, the larger the SAW amplitude $A_{\rm n}$, the faster the oil film moves due to the stronger induced leakage wave \citep{Shiokawa:1989tg}. Another notable aspect of SAW-induced dynamics is the contact line speed for early times ($< 1$ s): initially, it is rather small, but later increases and then remains constant. This change of contact line speed appears to be a consequence of the initial change of shape of the drop (see also Fig.~\ref{fig:exp_saw_intro}); while this is taking place, contact line speed is small.  After the initial deformation, the drop shape remains unchanged, translating approximately uniformly.
Figure~\ref{fig:exp_saw_results2}(b) shows consistent results for the maximum drop height, with an increase in height for early times, and constant values for longer times. Note that this uniform translation and constant film height differ significantly from the observed behavior for spreading under a body force such as gravity. These experimental observations will be explained and discussed within the framework of our theoretical model, which is described in the next section.  We note that in our experiments, we do not observe drop oscillations as in the experiments using water drops \citep{Brunet10}, possibly due to the much higher viscosity of the silicone oil.

\begin{figure}
\centering
\subfigure[]{
\includegraphics[width=0.485\linewidth]{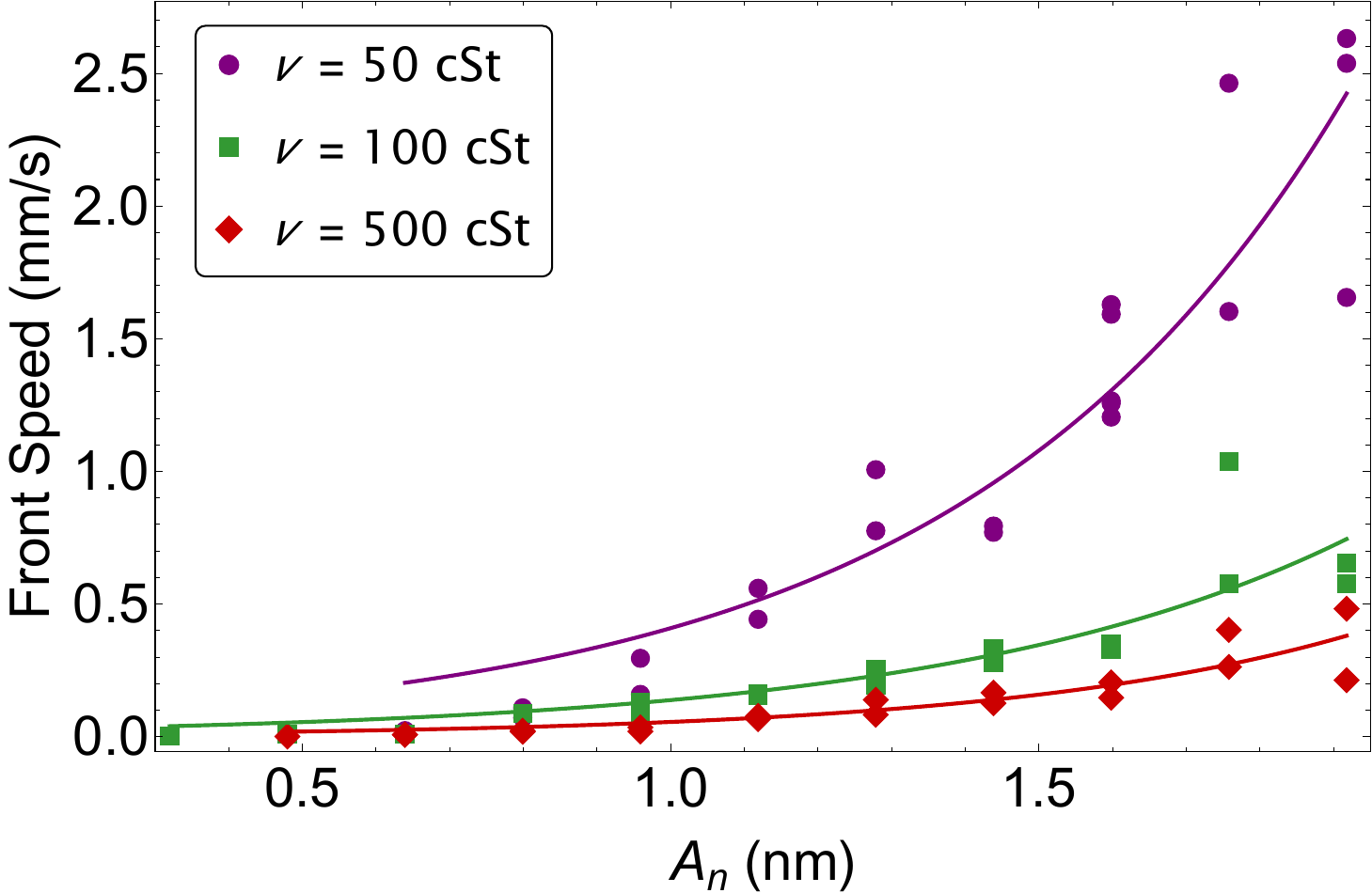}}
\subfigure[]{
\includegraphics[width=0.485\linewidth]{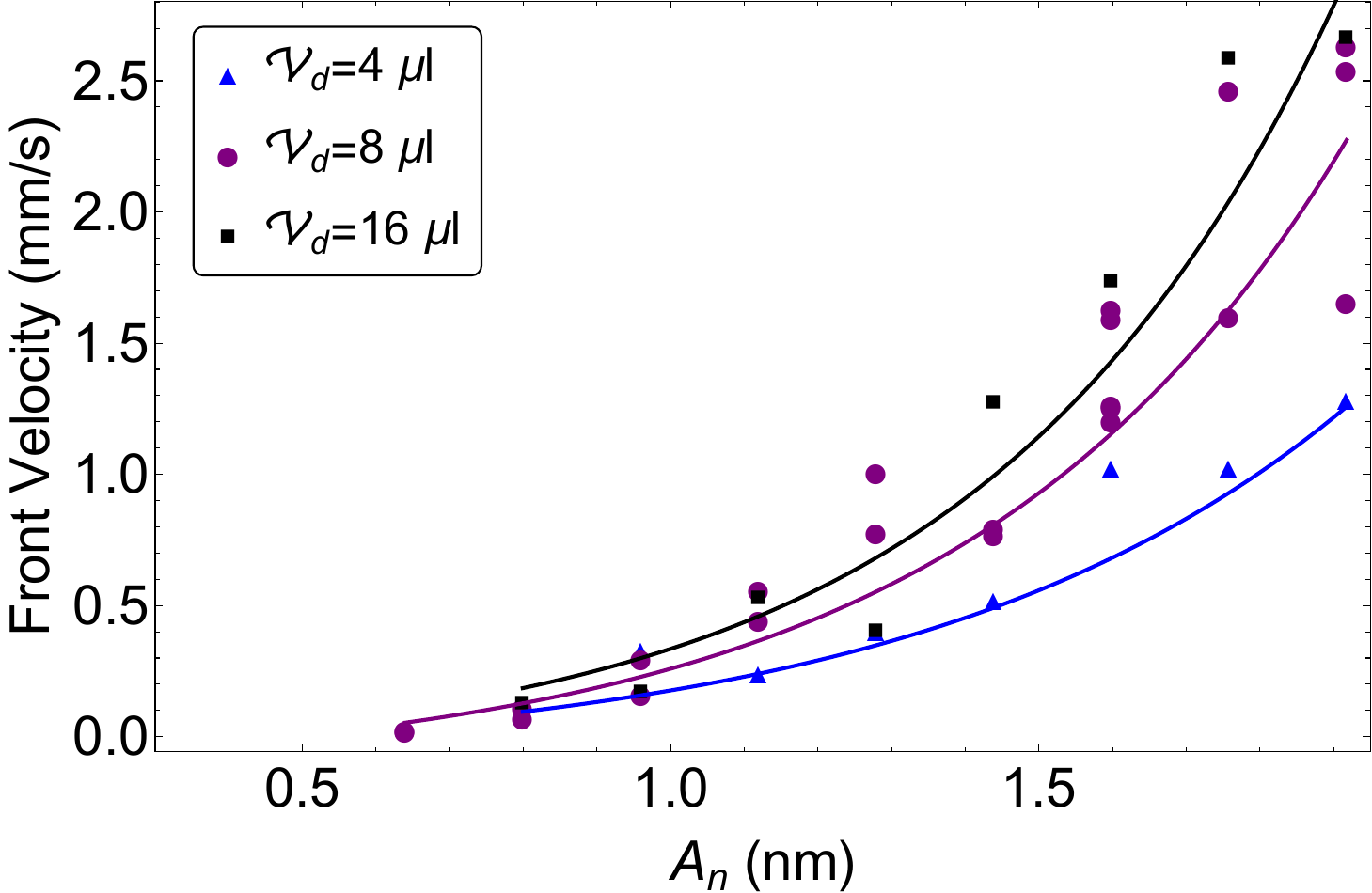}}
\caption{Asymptotic front speed of drop profiles for several values of the SAW amplitude $A_{\rm n}$ obtained using (a) silicone oil drops of volume $\mathcal{V}_{\rm d}=8\,\mu$l and different kinematic viscosities $\nu$; and (b) different volumes $\mathcal{V}_{\rm d}$ of silicone oil of the same viscosity $\nu=50$ cSt. We present two example videos comparing typical evolution at various kinematic viscosities and volumes in the supplementary materials (movie 2, movie 3).}
\label{fig:exp_saw_results3}
\end{figure}

In the experiments measuring the asymptotic front speed, we place a drop of silicone oil of various volumes ($4/8/16~\mu$l) and viscosities (50 cSt, 378356, Sigma-Aldrich; 100 cSt, 378364, Sigma-Aldrich; 500 cSt, 378380, Sigma-Aldrich) on the SAW actuator in a similar manner. We then apply different levels of SAW to induce the movement of the oil film. The initial frames of recorded side view videos are processed by public domain image processing software \emph{ImageJ} to mark a known distance of 5 mm (see the red lines in Fig.\ref{fig:exp_saw}(A-C) and movie 1, movie 2, movie 3 in the supplementary materials). The times at which the drop fronts pass these 5 mm markers are recorded. Since the marked area is positioned rather far from the IDT, we assume the drops have by this time settled to uniform translation, and we use these markers to calculate the average speed of the drops (of various volumes/viscosities); see  Fig.\ref{fig:exp_saw_results3}(a-b). Fig~\ref{fig:exp_saw_results3}(a) shows that front speed decreases as viscosity increases, assuming a fixed amplitude of the SAW. Interestingly, this decrease in front speed does \textit{not} scale linearly with viscosity (as is the case for drops under the influence of only gravity and capillarity, i.e. drops sliding down a ramp or spreading on a flat substrate under capillary forces) due to the nonlinear dependence of the acoustic terms on the viscosity, which is discussed in the model formulation in Sec.~\ref{sec:model}. Fig~\ref{fig:exp_saw_results3}(b) shows that the front speed increases as drop volume increases, an effect that is magnified for larger SAW amplitudes.

\section{Model}
\label{sec:model}

In our model development, we follow the approach of \citet{vanneste_streaming_2011} and \citet{campbell_propagation_1970}, implementing simplifications appropriate to our experiments.   One important point is that, for the considered parameters, the dominant contributions to the Reynolds stress (responsible for the force driving the flow) arise from both the SAW attenuation in the solid, and from viscous damping of the leaked wave. The latter effects were not included by~\citet{Shiokawa:1989tg}; we will comment in particular in Appendix~\ref{app:LeakySAW} on the consequences of this difference.

\begin{figure}
\centering
\includegraphics[width=0.7\linewidth]{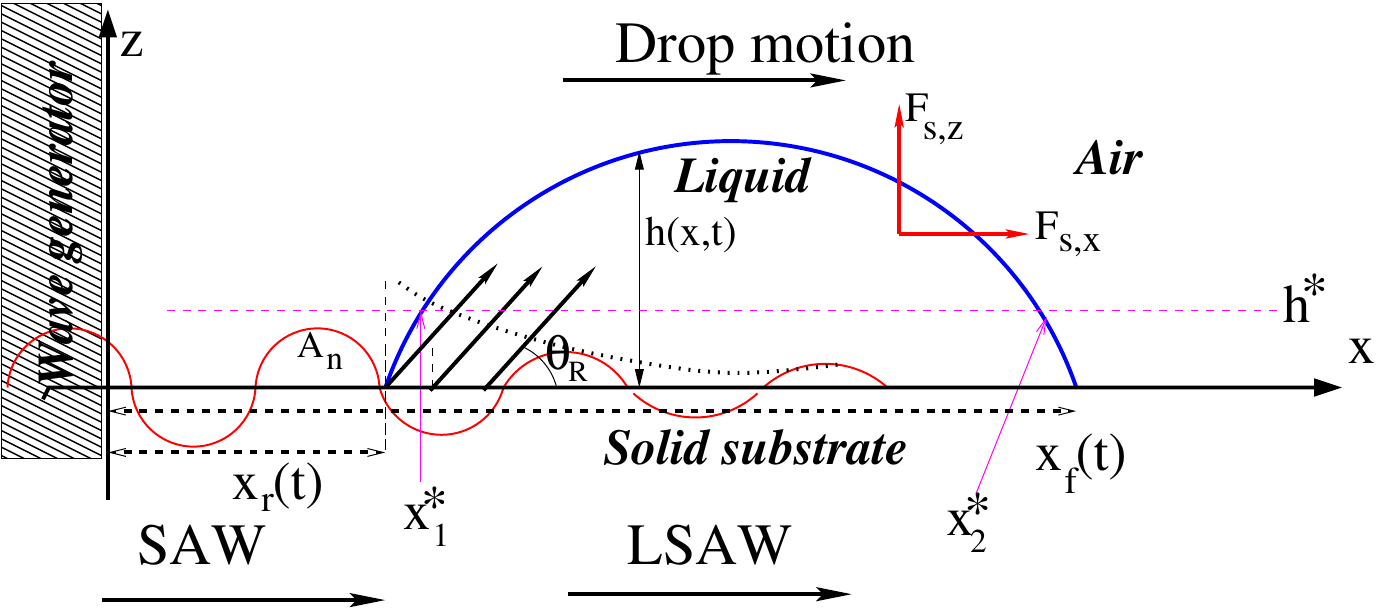}
\caption{Schematic of a liquid drop (blue curve) driven by a leaky SAW of amplitude $A_{\rm n}$. The SAW travels from left to right in the solid substrate.  The drop extends from the rear to the front contact lines at $x_{\rm r}(t)$ and $x_{\rm f}(t)$, respectively. The thin red line represents the amplitude $A_{\rm n}$ of the SAW at the substrate and shows how it is attenuated due to the presence of the liquid (the black dotted line is the envelope of the decaying amplitude in the liquid). The Rayleigh angle is denoted by $\theta_R$. The thickness $h^\ast$ and the positions $x^{\ast}_{1,2}$ are defined later in the text, see Eq.~(\ref{eq:phis_hcut}).}
\label{fig:scheme}
\end{figure}

We consider a two-dimensional Newtonian fluid layer occupying a domain in the $(x,z)$--plane, traversed by a sound wave generated by a Rayleigh-type-SAW traveling in a neighboring solid; see Fig.~\ref{fig:scheme} for a sketch. 
We assume that the associated acoustic field Mach number, $M_a \equiv\omega A_{\rm n}/c$  
(where $\omega$, $A_{\rm n}$ and $c$ are the SAW angular frequency, characteristic normal displacement amplitude 
at the solid surface, and the phase velocity of the acoustic field in the fluid, respectively) is small ($M_a \ll 1$), and that the velocity ($\mathbf{v}$), pressure ($p$), and density ($\rho$) fields in the fluid can be written as 
\begin{eqnarray}
\label{eq:V}
\mathbf{v} &=& \mathbf{v}_0+\mathbf{v}_1 + \mathbf{v}_2+\ldots = \mathbf{v}_0+M_a \,\bar{\mathbf{v}}_1 + M_a^2 \,\bar{\mathbf{v}}_2+\ldots, \nonumber \\
p &=& p_0 + p_1 + p_2+\ldots = p_0 + M_a \,\bar{p}_1 + M_a^2 \,\bar{p}_2+\ldots, \\ 
\rho &=& \rho_0 + \rho_1 +\rho_2+\ldots = \rho_0 + M_a \,\bar{\rho}_1 +M_a^2 \,\bar{\rho}_2+\ldots, \nonumber 
\end{eqnarray} 
where the leading order components of these fields are associated with quiescent liquid ($\mathbf{v}_0=0$, $p_0$ constant) of constant ambient density $\rho_0$; the first-order corrections ($\mathbf{v}_1,~p_1,~\rho_1$), represent an oscillatory flow associated with the acoustic field generated by the SAW that average to zero over the period of the acoustic forcing; and the second-order corrections ($\mathbf{v}_2,~p_2,~\rho_2$) represent the induced steady streaming flow. The assumption implicit in the above expansions in the Mach number is that the barred quantities are comparable in size to the corresponding leading-order contributions. The problem is governed by the Navier--Stokes (NS) and continuity equations, 
\begin{eqnarray}
\label{eq:NS1}
&{\partial (\rho \mathbf{v})}/{\partial t} + 
\rho \left( \mathbf{v} \cdot \nabla \right) \mathbf{v} +
\mathbf{v} \,\nabla \cdot \left( \rho \mathbf{v} \right) = 
-\nabla p + \mu \nabla^2 \mathbf{v} + 
\left( \mu_b +{\mu}/{3} \right) \nabla \left( \nabla \cdot \mathbf{v} \right) + \rho \mathbf{g}, \\ 
\label{eq:cont}
&{\partial \rho}/{\partial t}+\nabla\cdot\left(\rho \mathbf{v}\right)=0,
\end{eqnarray} 
where $t$ is time; $\mu$, $\mu_b$ are the shear and dilatational coefficients of the bulk viscosity, respectively; and  $\mathbf{g} =-g\mathbf{e}_z$ is the gravitational acceleration. In terms of the above expansions, we will assume that the gravitational term does not appear until second-order in $M_a$, which amounts to an assumption that $\mu v_1/\ell^2 \gg \rho g$ (where $v_1$ is a representative value of $|\mathbf{v}_1|$ and $\ell$ is a representative droplet size), ensuring that gravity is negligible in both leading-order and first-order problems. This condition was checked {\it a posteriori} and found to hold. With these assumptions, we substitute the expansions from Eq.~\eqref{eq:V} into Eq.~\eqref{eq:NS1} and collect terms of like orders in $M_a$; Secs.~\ref{sec:first} and~\ref{sec:second} discuss the first and second order terms, respectively.

\subsection{First order solution}
\label{sec:first}

The first order is associated with the propagation of acoustic waves in the fluid and hence we use the Helmholtz decomposition to separate the flow field into irrotational and divergence--free flow, i.e., $\mathbf{v}_1=\nabla \phi+\nabla\times\mathbf{\Psi}$, where $\phi$ is the velocity potential of the irrotational flow component, and $\mathbf{\Psi}$ represents the divergence--free flow component, so that the flow vorticity is equal to $-\nabla^2 \mathbf{\Psi}$. We ignore contributions from $\mathbf{\Psi}$, assuming $\mathbf{v}_1\approx \nabla \phi$~\footnote{The vorticity is known to be concentrated mainly in a boundary layer flow near the solid surface of thickness $\sqrt{\mu/(\rho_0\omega)}$, which amounts to just a few hundred nanometers and introduces only small contributions to the bulk flow far from the solid compared with the acoustic wave described by the potential $\phi$; see \citet{vanneste_streaming_2011} for a general approach that further accounts for the boundary layer flow.}. Here and in the following, it is understood that the real part of complex equations is taken when presenting results for real quantities such as $\phi$, ${\bf v}_1$ or ${\bf v}_2$. 
  
 For simplicity of notation we then set ${\bf v}_1 = \nabla \phi$ in Eq.~\eqref{eq:V}, leading (at order $M_a$ on substitution in Eqs.~\eqref{eq:NS1} and \eqref{eq:cont}) to the damped wave equation~\citep{vanneste_streaming_2011}, 
\begin{equation}
\label{eq:waveEq}
\phi_{tt}=c^2\nabla^2\phi+(\nu+\nu' )\nabla^2\phi_t,
\end{equation}
where $c^2=\left. dp/d\rho \right|_0$, $\nu'=\nu_b + \nu/3$, with $\nu\equiv\mu/\rho_0$ and $\nu_b\equiv\mu_b/\rho_0$. Assuming that the potential function, $\phi$, is harmonic in time with angular frequency $\omega$, and hence given by $\phi=\hat{\phi}({\bf x})\exp{(i\omega t)}$, Eq.~\eqref{eq:waveEq} may be reduced to 
\begin{equation}
\label{eq:helmholtz}
\nabla^2\hat{\phi}=- \frac{\omega^2}{c^2+i(\nu+\nu')\omega}\hat{\phi} = -\kappa^2 \hat{\phi},
\end{equation}
where
\begin{equation}
\kappa=\frac{k}{\sqrt{1+ i(\nu+\nu') k/c}}~, 
\label{eq:kappa}
\end{equation}
and $k= \omega/c=2\pi/\lambda$ is the wavenumber of the sound wave in the liquid, with wavelength $\lambda$. The real and imaginary parts of $\kappa$ are given by
\begin{equation}
\kappa= k_{\rm r} - i \,k_{\rm i}= k \sqrt{\cos \beta} \,e^{-i \beta/2} ,
\label{eq:kappa-def}
\end{equation}
assuming $\beta \in (0,\frac{\pi}{2})$ where
\begin{equation}
        \tan \beta = \left( \nu+\nu' \right) \frac{k}{c}.
        \label{eq:tan_beta}
\end{equation}

The solution of Eq.~\eqref{eq:helmholtz} requires a boundary condition that captures the corresponding leading order behavior for the normal component of a Rayleigh-type-SAW at the solid surface. This condition can be derived from the solution given by \citet{Royer1996} for the Rayleigh wave in an isotropic solid as 
\begin{equation}
\label{eq:bcmain}
\left. \frac{\partial \phi}{\partial z}\right|_{z=0}= A\omega\exp{(i\omega t)}\exp{(-i\kappa_{\rm s} x)},
\end{equation}
where subscript ${\rm s}$ stands for solid, and $A$ is the displacement amplitude in the liquid.
Unlike the solution in \cite{Royer1996}, the wavenumber $\kappa_{\rm s}$ has an imaginary part, i.e.,
\begin{equation}
\kappa_{\rm s}=k_{\rm s,{\rm r}}-i k_{\rm s,{\rm i}},  \label{eq:kappa_s}
\end{equation}
due to the presence of the liquid on top of the solid that converts the simple SAW into a leaky SAW (LSAW). 
The consequent attenuation of the LSAW is accounted for by the imaginary part of $\kappa_{\rm s}$, which is given by~\citep{Arzt:1967p112}
\begin{equation}
    \label{eq:ki}
    k_{\rm s,{\rm i}} =  \frac{\rho_0}{\rho_{\rm s}}\frac{c }{c_{\rm s}^2}f,
\end{equation}
where $\rho_{\rm s}$ is the solid density, $f=\omega/2\pi$ is the SAW frequency and $c_{\rm s}$ ($=\omega/k_{\rm s,{\rm r}}$) is the SAW phase velocity in the substrate. Note that the boundary condition, Eq.~\eqref{eq:bcmain}, is derived using the expansion of the nonlinear Rayleigh wave for small strains in the solid, so that the actual deformations at the solid surface as well as the inherently nonlinear relation between strain and stress in the solid, contribute at higher orders of $M_a$, not considered here. 

We comment here also on the relation between the experimentally-measured amplitude, $A_{\rm n}$, and the theoretical one, $A$, used in Eq.~(\ref{eq:bcmain}).  Although we expect $A\propto A_{\rm n}$, it is not clear that these amplitudes are identical, as discussed in some detail by \cite{Royer1996}. A detailed analysis to determine the exact relation between $A$ and $A_{\rm n}$ is beyond the scope of the present work, so we assume that $A = A_{\rm n}$, a point to which we return later when discussing the comparison between experimental and theoretical results.

\begin{table}
    \centering
    \begin{tabular}{|l|l|l|l|}\hline
        Physical parameters & Wave Parameters & Lengths in substrate & Lengths in oil\\ \hline
       $\rho_0 = 0.96$ g/cm$^3$& $f = 20$ MHz & $k_{\rm s,{\rm r}}^{-1} = 0.03$ mm  &$k_{\rm r}^{-1}= 0.01$ mm \\ 
       $\nu = \nu'= 50$ cSt & $\omega = 125.7$ MHz & $k_{\rm s,{\rm i}}^{-1} = 2.70$ mm  &$k_{\rm i}^{-1}= 2.56$ mm  \\ 
       $\gamma = 20.8$ dyn/cm & $A=1.04-10$ nm  &$\lambda_{\rm s} = 194$ $\mu$m & $\lambda = 67.5$ $\mu$m \\ 
       $c=1350$ m/s & $A_{\rm n}=0-1.69$ nm & $\delta= 0.91$ $\mu$m \\
       $\rho_{\rm s}=4.65$ g/cm$^3$ & & &  \\
       $c_{\rm s}=3880$ m/s & &  & \\ 
       $\ell = 1$ mm & & & \\
       \hline
    \end{tabular}
    \caption{Values of the physical parameters related to the experiments and the derived lengths for both the solid substrate and the liquid (PDMS).} 
    \label{tab:exper_data}
\end{table}

The experimental parameter values relevant to the silicone oil--lithium niobate system that we consider here are summarized in Table~\ref{tab:exper_data}. Note that, for the considered experimental parameters, we have $\tan \beta \approx \beta  \ll 1$ or $\nu+\nu' \ll  c/k$, so that to leading-order in $\beta$ (see Eqs.~(\ref{eq:kappa-def}, \ref{eq:tan_beta})),
\begin{equation}
   k_{\rm r} \approx k, \qquad k_{\rm i} \approx k \frac{\beta}{2} \approx \frac{k^2 \left( \nu+\nu' \right)}{2 c}.
   \label{eq:eta_zeta_approx}
\end{equation}
This approximate value of $k_{\rm i}$ is in agreement with that defined in Eq.~(5) of~\citet{Brunet10}. Note that $\beta \ll 1$ for a range of fluids; e.g., for water ($\nu=1$~cSt) and oil ($\nu=50$~cSt) we obtain $\beta=1.67\times 10^{-4}$ and $\beta=8.38\times 10^{-3}$, respectively, assuming $\nu_b=\nu$ in both cases. Therefore,  we can consider the ratio $k_{\rm i}/k_{\rm r}$ as a small parameter, as discussed further in  Appendix~\ref{app:LeakySAW}. 

The problem prescribed by Eqs.~\eqref{eq:helmholtz} and \eqref{eq:bcmain}, alongside a requirement that the acoustic field velocity decays far from the solid surface as $z\rightarrow \infty$, is satisfied by the potential function 
\begin{equation}
\label{HelmholtzEqSol}
\phi=\frac{- A\omega}{\sqrt{\kappa_{\rm s}^2-\kappa^2}}\exp({i\omega t})\exp{\left(-i\kappa_{\rm s} x\right)}
\exp{\left(-z\sqrt{\kappa_{\rm s}^2-\kappa^2} \right)},
\end{equation}
yielding the velocity components 
\begin{eqnarray}
\label{eq:velocity1}
v_{1,x}&=&\frac{\partial \phi}{\partial x}=\frac{i\kappa_{\rm s}}{\sqrt{\kappa_{\rm s}^2-\kappa^2}} A\omega\exp({i\omega t})
\exp{\left(-i\kappa_{\rm s} x\right)}\exp{\left(-z\sqrt{\kappa_{\rm s}^2-\kappa^2} \right)}, \nonumber \\
v_{1,z}&=&\frac{\partial \phi}{\partial z}=A\omega\exp({i\omega t})\exp{\left(-i\kappa_{\rm s} x\right)}
\exp{\left(-z\sqrt{\kappa_{\rm s}^2-\kappa^2} \right)}. 
\end{eqnarray}
For reference, the (physically relevant) real parts of these velocity components are written out explicitly in Appendix~\ref{app:RealVel}. 

We note that the analysis presented so far follows parts of the work by \citet{campbell_propagation_1970} and \citet{vanneste_streaming_2011}, albeit instead of calculating the SAW in the solid from conservation equations, we take the simpler approach of representing the component of the leaky SAW at the solid surface as a harmonic function that decays exponentially along its path \citep{Arzt:1967p112} and hence are able to obtain the analytical representation of the acoustic contribution to the velocity field in the fluid.  The formulation that ignores viscous dissipation is discussed in Appendix \ref{app:LeakySAW}.

\subsection{Second order solution}
\label{sec:second}
Having obtained $\mathbf{v}_1$, we next seek the quasi--steady flow field at times long compared to the acoustic period~\citep{Stokes1847,LordRayleigh1884, Schlichting:1932p447}. Since we excite the fluid using a single-frequency SAW, it is sufficient to time--average the system of equations over the period of the SAW~\citep{vanneste_streaming_2011} using the operator $\left<\cdot\right>\equiv \omega/(2\pi) \int_{t=0}^{2\pi/\omega}\cdot~dt$. The leading order velocity contribution that supports a steady flow component (and hence does not vanish) is $\left<\mathbf{v}_2\right>$, appearing at order $M_a^2$ in the expansions specified by Eq.~\eqref{eq:V}. The corresponding conservation of momentum and mass equations are then given by
\begin{eqnarray}
\label{eq:NS}
&-\nabla \left<p_2\right> + {\mathbf{F}}_{\rm s} -\rho_0 g\mathbf{e}_z + \mu \nabla^2 \left<\mathbf{v}_2\right>  +
\left( \mu_b + \frac{\mu}{3} \right) \nabla \left( \nabla \cdot \left<\mathbf{v}_2\right> \right)=0, \\ 
\label{eq:continuity}
&\rho_0\nabla\cdot\left<{\bf v}_2\right>+\nabla\cdot\left<\rho_1{\bf v}_1\right>=0,
\end{eqnarray}
where
\begin{equation}
\label{eq:Fs}
\mathbf{F}_{\rm s}= -\langle \rho_0 \left( \mathbf{v}_1 \cdot \nabla \right) \mathbf{v}_1 +
\mathbf{v}_1 \nabla \cdot \left( \rho_0 \mathbf{v}_1 \right) \rangle,
\end{equation}
represents the acoustic force on the induced steady streaming flow. Moreover, considering the component of the continuity equation proportional to $M_a$, $\partial \rho_1/\partial t+\rho_0\nabla\cdot{\bf v}_1=0$, and the solution for ${\bf v}_1$ specified by Eq.~\eqref{eq:velocity1}, we see that the product $\rho_1{\bf v}_1$ leads to an odd function over the SAW period, which results in $\left<\rho_1{\bf v}_1\right>=0$. Hence, Eqs.~\eqref{eq:NS} and \eqref{eq:continuity} may be simplified to
\begin{eqnarray}
\label{eq:NSv}
&-\nabla \left<p_2\right> + {\mathbf{F}}_{\rm s} -\rho_0 g\mathbf{e}_z + \mu \nabla^2 \left<\mathbf{v}_2\right>=0, \\
\label{eq:cont_v2}
&\nabla\cdot\left<{\bf v}_2\right>=0,
\end{eqnarray}
with ${\mathbf{F}}_{\rm s}$ as specified in \eqref{eq:Fs}. In Appendix \ref{app:RealVel} we use the real part of the first order velocity field, $\mathbf{v_1}$, to reveal the form of the SAW forcing as given in Eq.~\eqref{eq:Fs} above, and in Appendix~\ref{app:LeakySAW} we discuss a simplified version as well as the inviscid limit. The main points we emphasize are that the acoustic streaming force, given by Eq.~\eqref{eq:Fs}, is \textit{nonconservative} in contrast to the result in \cite{Shiokawa:1989tg}, and that its prefactor is significantly dependent on the presence of viscous dissipation. The discussion is made explicit in Appendix~\ref{app:RealVel} where the real component is shown along with the exact form of the coefficients therein.

\subsection{Long-wave approximation}

We carry out a standard long--wave expansion of Eqs.~(\ref{eq:NSv}) and (\ref{eq:cont_v2}), 
focusing on the simplified two-dimensional geometry that assumes
translational invariance in the transverse $y$-direction ($\partial /\partial y=0$). The discussion of three--dimensional effects will be presented elsewhere. Here, we obtain the pressure, $p \equiv \langle p_2\rangle$, by substituting the $z$-component of ${\mathbf{F}}_{\rm s}$, given in Eqs.~(\ref{eq:Fsx}) and (\ref{eq:Fsz}), in the $z$ component of Eq.~(\ref{eq:NSv}). We then integrate the resulting equation over the film thickness with respect to $z$, using the Laplace pressure boundary condition $p(x,h(x,t))=-\gamma h^{\prime \prime}$ (simplifying the curvature of the free surface in the spirit of the long--wave approximation, and using $\prime$ to denote $\partial/\partial x$), to find
\begin{equation}
    p(x,z) = -\gamma h^{\prime \prime} + \rho_0 g (h-z) +\frac{C_z P_0}{2K_z}\left[\psi(x,h)-\psi(x,z) \right] , \label{eq:pressure}
\end{equation}
where parameters $K_z, C_z$ are specified in Eqs.~\eqref{eq:kz},~\eqref{eq:Cz} respectively, and we defined the auxiliary function $\psi$ and parameter $P_0$ as
\begin{equation}
    \psi(x,z) = e^{-2(k_{{\rm s},{\rm i}}x+K_zz)}, \qquad P_0=\rho_0 A^2 \omega^2.
    \label{eq:psi}
\end{equation}
The last term on the right hand side of Eq.~(\ref{eq:pressure}) is the contribution of the bulk acoustic forcing to the local pressure in the film. To obtain the velocity component in the $x$--direction at this order, $v_{2,x}$, we substitute both the $x$--derivative of the pressure in Eq.~(\ref{eq:pressure}), and $F_{{\rm s},x}$ as given in Eq.~(\ref{eq:Fsx}), into the $x$--component of Eq.~(\ref{eq:NSv}). We then integrate twice with respect to $z$, applying a zero shear-stress boundary condition at the free surface $(\partial v_{2,x}/\partial z=0$ at $z=h$) and a no-slip condition at the substrate $(v_{2,x}=0$ at $z=0)$, which yields 
\begin{align}
    v_{2,x} = &-\frac{1}{2\mu} z(z-2h)\left[\gamma h^{\prime \prime \prime}-\rho_0 g h^{\prime} + C_z P_0\psi(x,h)\left(\frac{k_{\rm s,{\rm i}}}{K_z}+h^{\prime}\right) \right] \nonumber \\
    &\quad - \mathcal{C} \frac{P_0}{4\mu K_z^3}e^{-2k_{\rm s,{\rm i}} x}\left(e^{-2K_zz}+2K_ze^{-2K_zh}z-1 \right) , \label{eq:Vel2}
\end{align}
where 
\begin{equation}
    \mathcal{C} = K_z C_x -k_{\rm s, i} C_z , \label{eq:scriptC}
\end{equation}
with parameter $C_x$ defined in~\eqref{eq:Cx}.  
Note that the last term in Eq.~(\ref{eq:Vel2})
leads to a velocity profile $v_{2,x}$ that is modified from the usual parabolic shape in the neighborhood of $z=0$. Interestingly, there is a critical value of $\mathcal{C}$, namely $\mathcal{C}_{\rm crit}<0$, such that for $\mathcal{C}<\mathcal{C}_{\rm crit}$, the velocity profile includes negative values of $v_{2,x}$ in the region near the substrate. We have not, however, observed such flow inversion with the present choice of parameters relevant to our experiments.

From the velocity profile above, Eq.~(\ref{eq:Vel2}), we obtain the film-averaged velocity and flux as
\begin{equation}
    u = \frac{1}{h}\int\limits_0^h v_{2,x} \ dz, \qquad Q=uh.
\end{equation}
Conservation of mass for the fluid then requires that
\begin{equation} 
    \label{eq:dhdtdim}
    \frac{\partial h}{\partial t} + \frac{\partial Q}{\partial x}=0.
\end{equation}
Upon using the velocity profile as obtained in Eq.~(\ref{eq:Vel2}) above, we find
\begin{equation}
    Q = -\frac{h^3}{3\mu}\frac{\partial {\mathcal P}}{\partial x} - \mathcal{C}\frac{P_0}{8\mu K_z^4}\psi(x,h)\left[ 2K_z^2h^2-1+e^{2K_zh}\left( 1-2K_zh \right)\right],
    \label{eq:Q_dim}
\end{equation}
where we have defined the effective pressure, 
\begin{equation}
    {\mathcal P} = -\gamma h^{\prime \prime} + \rho_0 gh+\frac{C_z P_0}{2K_z}\psi(x,h). \label{eq:P2}
\end{equation}
The first term on the right hand side of $\mathcal{P}$ is the capillary contribution to the quasi--steady pressure in the fluid film, wherein $\gamma$ is the surface tension at the fluid--air interface, assumed constant; the second and third terms account for the gravitational body force; and the final term models the contribution from the acoustic force due to the SAW. 

For our simulations, we nondimensionalize the problem using an arbitrary length scale, $\ell$, a timescale $t_c=3\mu \ell/\gamma$ such that leading order terms balance in Eq.~\eqref{eq:dhdtdim}, and a pressure scale $p_{\rm c} =\gamma/\ell$ based on the capillary contribution in Eq.~\eqref{eq:P2}, so that we have
\begin{equation}
(x,h)=\ell (\tilde x,\tilde h), \quad  t =\frac{3\mu \ell}{\gamma} \tilde t, \quad {\mathcal P} = \frac{\gamma}{\ell} \tilde{{\mathcal P}}, \quad (k_{\rm s,{\rm i}},K_z,C_x,C_z)=\ell^{-1} (\tilde{k}_{\rm s,{\rm i}},\tilde{K}_z,\tilde{C}_x,\tilde{C}_z), \quad {\cal C}=\ell^{-2} \tilde{\cal C}.
\label{eq:scalings}
\end{equation}
Therefore, in dimensionless form, we have 
\begin{equation}
    \frac{\partial \tilde{h}}{\partial \tilde{t}}+\frac{\partial}{\partial \tilde{x}}\bigg[-\tilde{h}^3 \frac{\partial \tilde{{\mathcal P}}}{\partial \tilde{x}} - \tilde{\mathcal{C}} \frac{3\mathcal{S}}{8 \tilde{K}_z^4}\psi(\tilde{x},\tilde{h})\Big(2 \tilde{K}_z^2 \tilde{h}^2-1+e^{2 \tilde{K}_z \tilde{h}}(1-2 \tilde{K}_z \tilde{h}) \Big) \bigg]=0, \label{eq:dhdtadim}
\end{equation}
where it is understood that in calculating $\psi(\tilde{x},\tilde{h})$ we use Eq.~\eqref{eq:psi} with $k_{\rm s,\rm i}$, $K_z$ replaced by their tilded equivalents; the dimensionless effective pressure is
\begin{equation}
    \tilde{\mathcal P} = -\tilde{h}^{\prime \prime} + {\rm Bo} \tilde{h}  + \frac{\mathcal{S} \tilde{C}_z}{2 \tilde{K}_z} \psi( \tilde{x},\tilde{h}),
    \label{eq:Pad}
\end{equation}
and we have defined the nondimensional parameters
\begin{equation}
    {\rm Bo}=\frac{\rho_0 g \ell^2}{\gamma}=\frac{\ell^2}{a^2}, \quad \mathcal{S} = \frac{P_0 \ell}{\gamma} = \frac{\rho_0 \ell A^2 \omega^2}{\gamma}, \label{eq:Bo-S}
\end{equation}
where $a=\sqrt{\gamma/(\rho_0 g)}$ is the capillary length. In our simulations, motivated by a typical drop size in the experiments, we set $\ell = 1$ mm. Table \ref{tab:param} provides a list of nondimensional parameter values used in all simulations unless otherwise stated.

We pause here to provide a justification for neglecting the effects of acoustic radiation pressure in the present model. The experiments modeled here involve a SAW propagating in the solid with attenuation length of $1$-$2$ mm, while the drops considered are $6$-$7$ mm in lateral length. When comparing the relative strengths of acoustic streaming and acoustic radiation pressure, an important distinction is that streaming results from the SAW in the solid, while acoustic radiation pressure results from the leaky SAW in the liquid that reflects off the drop's free surface. The attenuation lengths of both effects are proportional to the relative phase velocities of the acoustic fields in the solid and fluid, respectively. Since the speed of sound in the liquid is $2$-$3$ times smaller than the speed of sound in the solid, the radiation pressure attenuates twice as fast as the SAW in the solid. Hence, the radiation pressure affects the drop rear only along a distance of approximately $0.5$-$1$ mm. Because the present focus is on describing the front dynamics of \textit{driven} drops, we neglect this contribution. However, we note that in cases where the attenuation length of the SAW is comparable to the lateral length of the drop, acoustic radiation pressure is expected to have an effect on the drop geometry and dynamics comparable to that of acoustic streaming, and hence should be included (see,~e.g.,~\cite{marcos2025Monte}). To quantify the regime in which radiation pressure becomes important, some insight can be reached by considering the nondimensional ratio
\begin{equation}
    R=\frac{1}{2r_{\rm d}k_{\rm i}}, 
\end{equation}
that is, the attenuation length of the SAW in the solid divided by the lateral length of the drop. When $R = \mathcal O (1)$, acoustic radiation pressure should be considered to properly model the drop front, but when $R\ll 1$ (e.g., $R=1/5$ as in our case), it can be neglected when modeling the drop front dynamics.

\section{Theoretical results}
\label{sec:results}

In this section, we discuss the predictions of the model developed in Sec.~\ref{sec:model}, and compare them with the experimental results.  In Sec.~\ref{sec:basic_case}, we present results for a few chosen reference cases, and then we discuss the influence of model parameters, including the acoustic amplitude, $A$, the droplet volume $\mathcal{V}_{\rm d}$ and the droplet kinematic viscosity $\nu$, on the results.  In Sec.~\ref{sec:travel} we discuss (approximate) traveling-wave solutions. 

\subsection{Main features of the results}
\label{sec:basic_case}

In our simulations, we solve Eq.~(\ref{eq:dhdtadim}) numerically using COMSOL\texttrademark; see Appendix~\ref{app:comsol} for the description of the implementation.  As the initial condition, we consider the two--dimensional parabolic drop of (dimensionless) cross-sectional area 
\begin{equation}
\label{eq:Ad}
\tilde{{\cal A}}_{\rm d} = 2\int_0^{\tilde{r}_{\rm d}} \tilde{h}_{\rm d}  \left[ 1- \left(\frac{\tilde{x}}{\tilde{r}_{\rm d}}\right)^2 \right] \,d \tilde{x} =
\frac{4}{3} \tilde{h}_{\rm d}  \tilde{r}_{\rm d}, 
\end{equation}
(recall that tilded quantities are dimensionless), which remains constant during the drop evolution.  The drop height parameter $\tilde{h}_{\rm d}$ is obtained from the known experimental (3D) drop volume, ${\cal V}_{\rm d} =\ell^3 \tilde{\cal V}_{\rm d}$, as $\tilde{h}_{\rm d} = 2 \tilde{\cal V}_{\rm d}/(\pi \tilde{r}_{\rm d}^2)$, using the formula for the volume $\tilde{\cal V}_{\rm d}$ of a 3D parabolic cap of base radius $\tilde{r}_{\rm d}$ and height $\tilde{h}_{\rm d}$. The 2D parabolic initial condition is centered at $\tilde{x}=\tilde{x}_{\rm d}$, and is written as
\begin{equation}
        \label{eq:hinit2D}
   \tilde{h}(\tilde{x},0)= \left\{ 
        \begin{array}{ll}
             \tilde{h}_{\rm p} ,&  \tilde{x} <  \tilde{x}_{\rm r}(0) ,\\
            \left( \tilde{h}_{\rm d} -  \tilde{h}_{\rm p} \right) \left[ 1- \left(\frac{ \tilde{x}- \tilde{x}_{{\rm d}}}{ \tilde{r}_{\rm d}}\right)^2 \right] +  \tilde{h}_{\rm p} ,& 
            \tilde{x}_{\rm r}(0)\leq  \tilde{x} \leq  \tilde{x}_{\rm f}(0),\\ 
             \tilde{h}_{\rm p} ,& \tilde{x} >   \tilde{x}_{\rm f}(0), 
        \end{array} 
        \right.
\end{equation}
where $\tilde{x}_{\rm r}(0)= \tilde{x}_{\rm d} - \tilde{r}_{\rm d}$ and  $ \tilde{x}_{\rm f}(0)= \tilde{x}_{\rm d} +  \tilde{r}_{\rm d}$ are the initial positions of the rear and front contact lines of the drop, respectively (see Fig.~\ref{fig:scheme}).  Here, $ \tilde{h}_{\rm p}$ ($ \ll  \tilde{h}_{\rm d}$) is the thickness of a precursor film, introduced to avoid the stress singularity associated with a moving contact line.  It is well known (see, e.g.~\cite{DK_jcp02}), that the step size $\Delta \tilde{x}$ in the spatial discretization should be similar to $\tilde{h}_{\rm p}$ to ensure numerical convergence, and therefore we use $\Delta \tilde{x} = \tilde{h}_{\rm p}$ in our simulations. The specific value assigned to $\tilde{h}_{\rm p}$ has only a minor influence on the results; e.g., if $\tilde{h}_{\rm p}$ is doubled or halved, the change of spreading speed or drop height is just 1-2\%.
All reported results are obtained using $\tilde{h}_{\rm p} = 0.01$.

In implementing our model for the geometry defined above, we note that in the experiments, the SAW experiences minimal attenuation in regions of the actuator where fluid is not present, with significant attenuation only under the bulk droplet. To model this, we modify the factor $\psi$ defined in Eq.~\eqref{eq:psi}, setting it to a constant value (no attenuation) in regions where the film height is smaller than some critical thickness $\tilde{h}^*$, with the exponential attenuation only where $\tilde{h}>\tilde{h}^*$, so that
\begin{equation}
    \psi(\tilde{x},\tilde{h}) = \left\{ 
    \begin{array}{ll}
    1, & \tilde{x}< \tilde{x}_1^\ast(\tilde{t})\, , \\
    e^{-2 \left[ \tilde{k}_{{\rm s},{\rm i}} (\tilde{x} - \tilde{x}_1^\ast(\tilde{t})) + \tilde{K}_z  \left( \tilde{h} - \tilde{h}^\ast \right) \right]}, & \tilde{x}_1^\ast(\tilde{t}) \leq \tilde{x} \leq \tilde{x}_2^\ast(\tilde{t})\, , \\
    e^{-2 \tilde{k}_{{\rm s},{\rm i}} \left[ \tilde{x}_2^\ast(\tilde{t}) - \tilde{x}_1^\ast(\tilde{t}) \right]}, & \tilde{x} > \tilde{x}_2^\ast(\tilde{t}) \, ,
    \end{array}
    \right.
    \label{eq:phis_hcut}
\end{equation}
where $\tilde{x}_1^\ast$, $\tilde{x}_2^\ast$ are defined by $\tilde{h}(\tilde{x}_1^\ast,\tilde{t})=\tilde{h}(\tilde{x}_2^\ast,\tilde{t})=\tilde{h}^\ast $, and the SAW is assumed to propagate from $\tilde{x}< \tilde{x}_1^*(\tilde{t})$ in the direction of increasing $\tilde{x}$.
In line with our discussion in Sec.~\ref{sec:model}, in simulations we choose the cutoff film thickness $\tilde{h}^*$ based on the thickness below which the bulk acoustic streaming becomes weak and mass transport is governed by other mechanisms (e.g., Rayleigh streaming and acoustic pressure~\citep{Rezk12,Rezk14}), $\tilde{h}^\ast \approx (\lambda_{\rm oil}/4)/\ell =17 \mu {\rm m}/\ell $.  

\begin{figure}
\centering
\subfigure[$A=1.44$ nm]{
\includegraphics[width=0.48\linewidth]{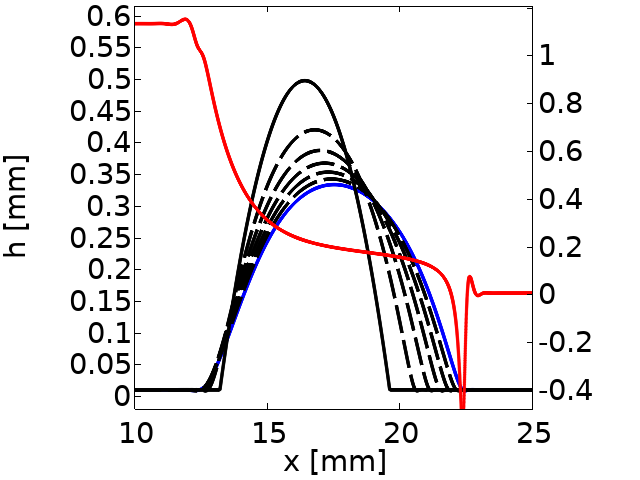}}
\subfigure[$A=2$ nm]{
\includegraphics[width=0.48\linewidth]{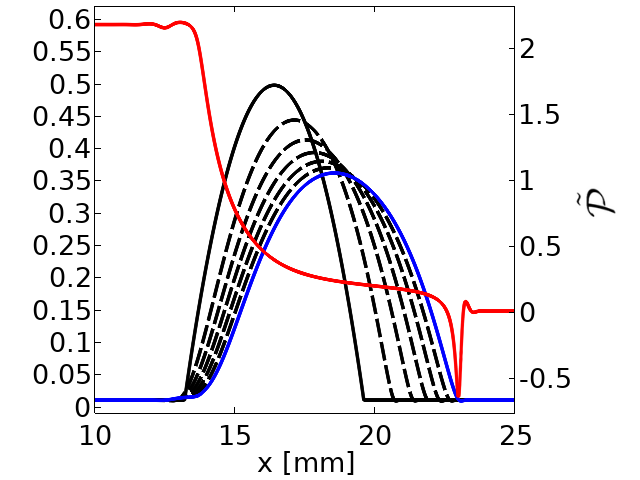}}
\subfigure[$A=6$ nm]{
\includegraphics[width=0.48\linewidth]{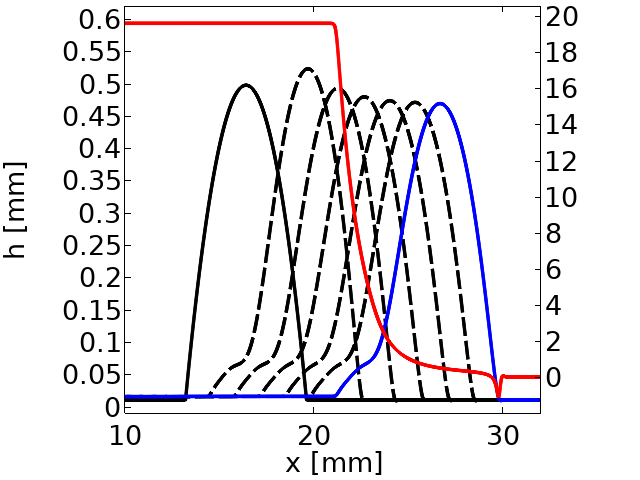}}
\subfigure[$A=10$ nm]{
\includegraphics[width=0.48\linewidth]{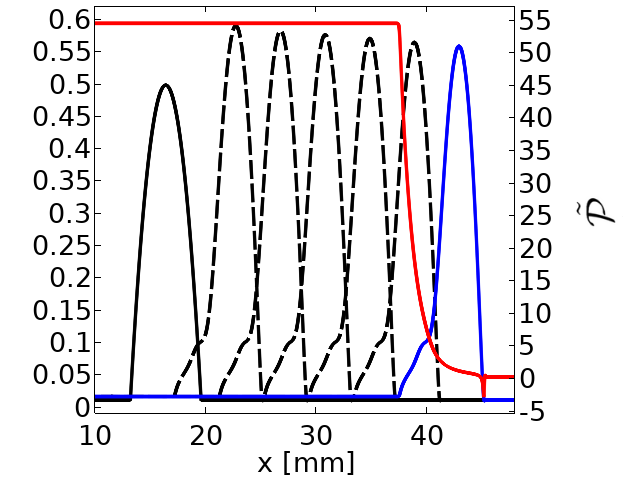}}
\caption{Evolution of thickness profile for different values of $A$. The black solid lines correspond to the initial condition, the dashed black lines to the profiles every time step $\Delta t=1$~s, and the solid blue line to $t=6~$s. The red line stands for the effective dimensionless pressure $\tilde{\mathcal{P}}$ at $t=6$~s.  Note that the dip in the pressure curve corresponds to the dramatic change of curvature of the fluid interface at the front contact line. }
\label{fig:hx_num}
\end{figure}

We commence our analysis by considering the experimental case where an oil drop of volume $8\, \mu$l is under the excitation of the $20$ MHz SAW. 
\begin{table}
    \centering
    \begin{tabular}{|l|l|l|l|} \hline
         Parameter & Definition & Numerical Value & Physical Meaning \\ \hline
         $\mathcal{S}$ & $(\rho_0\ell A^2 \omega^2)/\gamma$  & $0.788-72.88$ & Ratio of acoustic force and surface tension\\
         ${\rm Bo}$ & $(\rho_0 g \ell^2)/\gamma$ & $0.452$ & Bond Number \\
         $\tilde{\mathcal{C}}$ & $\tilde{K}_z \tilde{C}_x -\tilde{k}_{\rm s,{\rm i}} \tilde{C}_z$ & $-0.11$ & Effect of non-conservative SAW force \\ 
         $\tilde{K}_z$ & See Eqs.~\eqref{eq:scalings},~\eqref{eq:kz} & $0.279$ & Attenuation coefficient of SAW in $z$ \\
         $\tilde{h}^\ast$ & $(\lambda/4)/\ell$ & $0.017$ & Cutoff Thickness at which attenuation begins \\
         $\tilde{C}_z$ & See Eqs.~\eqref{eq:scalings},~\eqref{eq:Cz} & $0.415$ & Coefficient of $z$ component of SAW Force \\
         $\tilde{\mathcal{A}}_{\rm d}$ & $(4/3)\tilde{h}_{\rm d} \tilde{r}_{\rm d}$ & $2.12$ & Area of drop \\
         $\tilde{h}_{\rm d}$ & $(2 \tilde{\mathcal{V}}_{\rm d})/(\pi \tilde{r}_{\rm d}^2)$ & $0.497$ & Height of drop \\
         $\tilde{r}_{\rm d}$ & & $3.2$ & Radius of drop \\
         $\tilde{h}_{\rm p}$ & & $0.01$ & Precursor film thickness \\
         \hline
    \end{tabular}
    \caption{Dimensionless parameters used in the simulation of the experiment for a drop volume of $\mathcal{V}_{\rm d}=8\, \mu$l  driven by a SAW with $A$ in the range from Table \ref{tab:exper_data}, which determines the range of $\mathcal{S}$.}
    \label{tab:param}
\end{table}
Figure~\ref{fig:hx_num} shows the evolution of the thickness profile obtained by solving Eq.~(\ref{eq:dhdtadim}) for $A=1.44, 2, 6, 10$~nm  
subject to the initial condition given by Eq.~(\ref{eq:hinit2D}), and boundary conditions fixing the film thickness to $h_{\rm p}$ and requiring vanishing derivatives at the domain boundaries; the other parameter values are shown in Table \ref{tab:param}. For the low-amplitude cases (e.g. $A<2$~ nm, Fig.~\ref{fig:hx_num}(a) and (b)), we see that capillary and gravitational forces dominate, leading to a continual spreading of the drop. The acoustic force is able to slightly drive the drop along the direction of propagation of the SAW, as is apparent by the movement of the two contact lines. In contrast, in the high-amplitude cases (e.g., $A>2$~nm, Fig.~\ref{fig:hx_num}(c) and (d)), the acoustic force plays a dominant role in governing the dynamics. We see that the evolution of the drop can be described by two regimes: an initial transient stage where the drop takes on a new shape, followed by a period of translation across the domain at near-constant speed, consistent with experimental results (see Fig.~\ref{fig:exp_saw_results2}); as $A$ increases, the duration of the transient stage decreases. The fluid region near the rear contact line, where the pressure gradient is largest, serves as a ``snow plow" that pushes the bulk drop along the direction of propagation of the SAW. 
The result is a near traveling wave solution at long times, discussed further in Sec.~\ref{sec:travel}. The subtle, but continuous, decrease of the maximum height of the drop during this period is due to a thin trailing film (of approximate thickness $h^\ast$) that is left behind the bulk drop. We also note formation of a trailing foot forming spontaneously behind the main body of the drop for amplitudes $A> 2$~nm. This feature travels together with the main body of the droplet, in contrast to the trailing film, which stays behind. We direct the interested reader to the supplementary materials where animations for each of these four cases are shown (movie 4, movie 5, movie 6, movie 7).

\begin{figure}
\centering
\subfigure[]{
\includegraphics[width=0.48\linewidth]{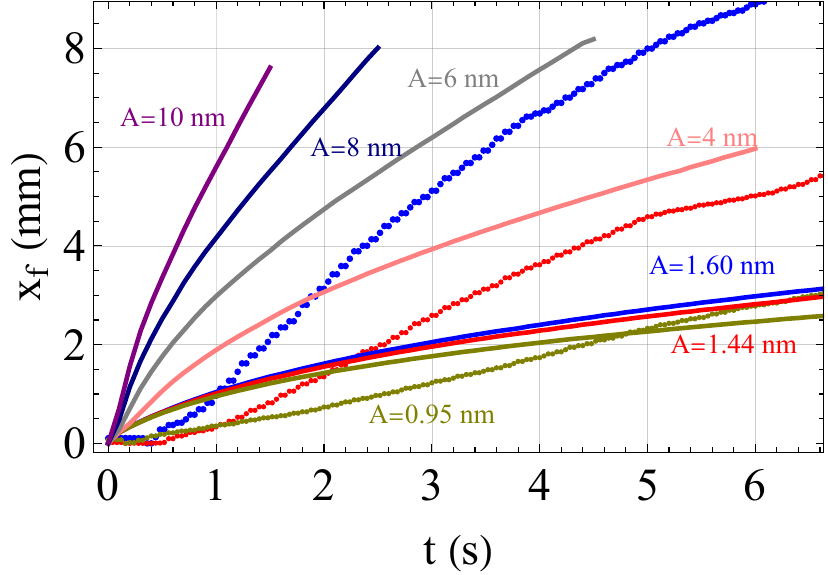}}
\subfigure[]{
\includegraphics[width=0.48\linewidth]{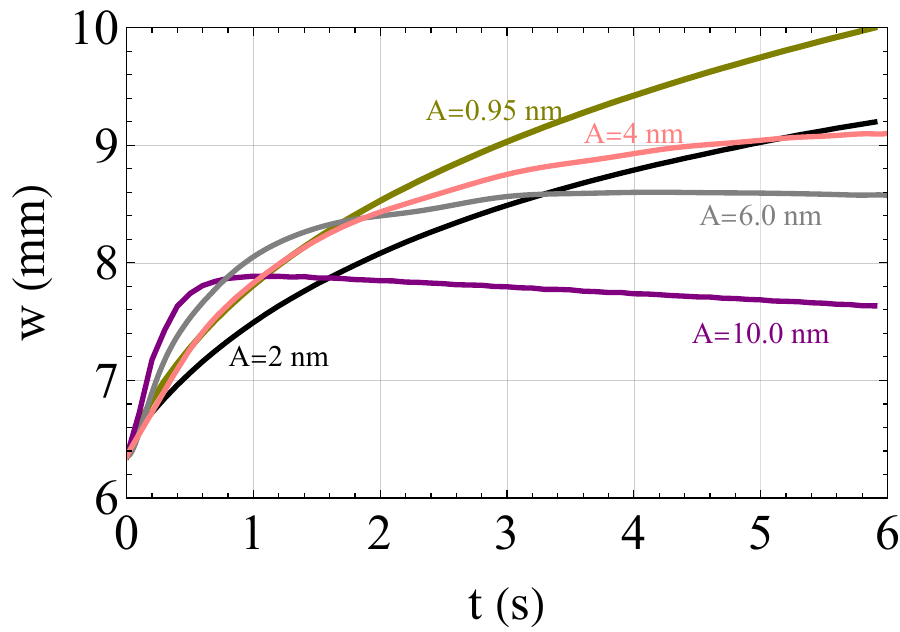}}
\caption{(a) Comparison between numerical (solid lines, amplitude values $A$ as labeled, color-coded) and experimental results (solid dots, for measured amplitude values $A_{\rm n}=1.60\,$nm ({\color{blue}blue}), $1.44\,$nm ({\color{red}red)}, $0.95\,$nm ({\color{olive} olive})). (b) Numerically-calculated drop width data plotted as a function of time for several $A$-values (color-matched with panel (a) where appropriate).}
\label{fig:As}
\end{figure}

Focusing next on the spreading speed, Fig.~\ref{fig:As} plots the relative speed of spreading for both experiments and simulations. The numerical results (solid lines) show a monotonic increase in spreading speed as $A$ increases, consistent with the experimental results (symbols). Note the distinct regimes in which the drops evolve, evident from the apparent change of slope of the curves, especially prominent in large-amplitude cases. Figure~\ref{fig:As}(b) shows the numerically-calculated drop width, ${\rm w}=x_2^\ast-x_1^\ast$, as a function of time, for different values of $A$. We once again emphasize the different behaviors exhibited by low- and high-amplitude solutions. For low-amplitude cases, the front contact line of the drop moves faster than the rear contact line, causing the drop to become thinner and wider. Conversely, for the high-amplitude cases, $\rm{w}$ initially increases and then remains nearly constant, or even decreases slowly.   

Figure~\ref{fig:As}(a) illustrates a quantitative difference between experiment and simulations: assuming that  $A = A_{\rm n}$ (see also the relevant discussion in Sec.~\ref{sec:model}), the simulated drops travel at a slower pace.  While it is not yet clear why this difference arises, some plausible reasons, in addition to possible differences between $A$ and $A_{\rm n}$~\citep{Royer1996}, include: (i) the potential relevance of inertial effects that are not included in the model at order $M_a$ (the solution $\mathbf{v}_1$)~\citep{ZAREMBO1971,Orosco2021UnravelingTC,Dubrovski_Friend_Manor_2023}; (ii) the omission of contributions from the boundary layer flow near the solid surface, i.e., the Rayleigh law of streaming; (iii) the capillary waves that appear at the free surface of the film \citep{LIGHTHILL:1978p12,Morozov17}; (iv) the use of a two-dimensional model to understand a three-dimensional system; and (v) the neglect of the ultrasonic wave reflection off the free surface of the liquid.  We have verified that the numerical results are independent of the precise values assigned to $h^*$ and $h_{\rm p}$, suggesting that neither boundary layer thickness nor precursor thickness are relevant here.  We have also carried out some preliminary simulations of a three-dimensional oil drop, which suggest that the two-dimensional model used in the present work is not responsible for the quantitative differences between experiments and simulations (we leave the discussion of three-dimensional effects for future work). Clearly, more work will be needed to uncover the sources of the quantitative differences between experiments and simulations.

\begin{figure}
\centering
\includegraphics[width=0.6\linewidth]{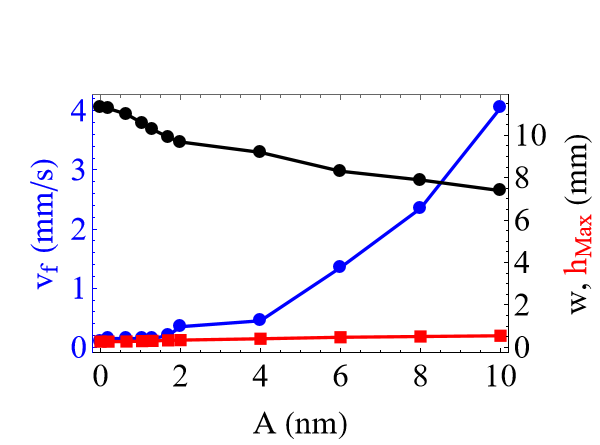}
\caption{Long time (calculated at $t=9$~s) values of the front speed, $\rm{v}_{\rm f}$ (in mm/s, {\color{blue}blue} filled circles), maximum thickness, $h_{\rm Max}$ (mm, {\color{red}red} filled squares), and drop width, $\rm w$ (cm, {\color{black}black} filled circles), as functions of $A$. Note that $\rm v_{\rm f}$ is plotted using the left $y$-axis while $\rm w$ and $h_{\rm {Max}}$ are plotted using the right $y$-axis. The points correspond to the raw data, and the solid lines connecting them guide the eye.}
\label{fig:width}
\end{figure}

To summarize the numerical results presented so far, Fig.~\ref{fig:width} plots the numerically-predicted values of the front speed, ${\rm v}_{\rm f}$, the maximum drop height, $h_{\rm Max}$, and the drop width, $\rm{w}$, all calculated at $t=9$~s, as a function of SAW amplitude $A$. This figure illustrates the transition between the two types of dynamics observed: for low and moderate $A$-values ($A\leq 2$ nm) we observe near-symmetric drop spreading due primarily to capillary forces, however, for large values ($A>2$ nm) the front speed increases approximately quadratically as $A$ increases, while the maximum height/width of the drop increases/decreases approximately linearly with respect to the amplitude value.

\begin{figure}
\centering
\subfigure[$\nu=\nu'=50\, \text{cSt}$]{
\includegraphics[width=0.32\linewidth]{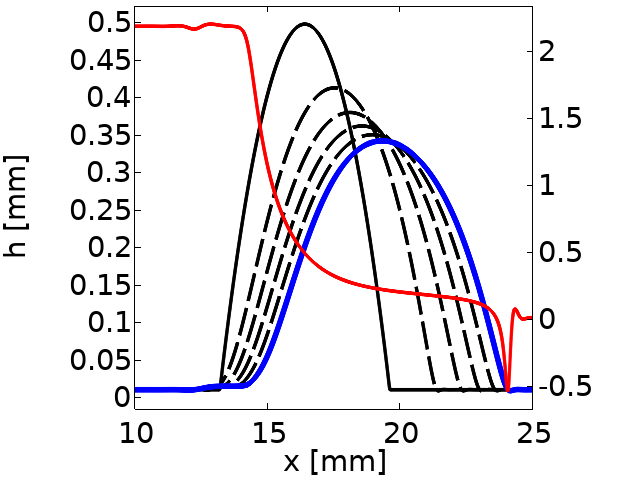}}
\subfigure[$\nu=\nu'=100\, \text{cSt}$]{
\includegraphics[width=0.32\linewidth]{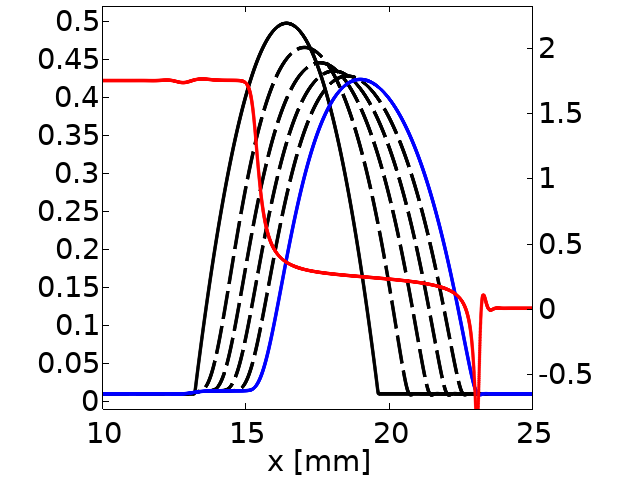}}
\subfigure[$\nu=\nu'=500\, \text{cSt}$]{
\includegraphics[width=0.32\linewidth]{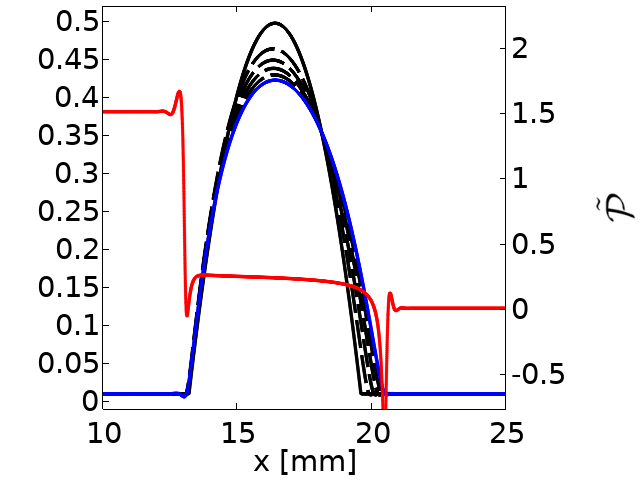}}
\caption{Evolution of thickness profile for $A=2$ nm for drops of different viscosities, $\nu=\nu'=50,100,500$ cSt, indicated in each figure part. The black solid lines correspond to the initial condition, the dashed black lines to the profiles every time step $\Delta t=2$~s, and the solid blue line to $t=10~$s. The red line denotes the effective dimensionless pressure $\tilde{\mathcal{P}}$ at $t=10$~s.}
\label{fig:diffVisc}
\end{figure}

Before concluding this section, we briefly discuss how the results depend on the oil viscosity. We have also carried out simulations using different drop volumes; while the results are consistent with the experimental ones, shown in Fig.~\ref{fig:exp_saw_results3}b, the influence of the change of fluid volume is, however, weaker than in the experiments (the results not shown for brevity).  When varying the kinematic viscosity, $\nu$ (assuming $\nu'=\nu$), without acoustic forcing ($A=0$), viscosity only (linearly) affects the time scale. With acoustics, however, the dynamics change in a nonlinear fashion, see Fig.~\ref{fig:diffVisc}. The mechanism behind this nonlinear dependence is traced to the prefactor of the nonconservative term in Eq.~\eqref{eq:dhdtadim}, which decreases as $\tilde{K}_z$ (defined in Eq.~\eqref{eq:kz}) grows and $\tilde{\mathcal{C}}$ (defined in Eq.~\eqref{eq:scriptC}) decreases in magnitude, yielding a relatively large attenuation length. This substantially weakens the forcing at the front, while nearly balancing gravitational and capillary forces at the rear, effectively stalling the drop motion for large viscosity values, in qualitative agreement with the experimental findings of Fig.~\ref{fig:exp_saw_results3}(a).

\subsection{Traveling wave solution}
\label{sec:travel}

In view of the results presented so far, we now briefly explore a possible traveling wave solution of Eq.~(\ref{eq:dhdtadim}) which appears present for sufficiently large SAW amplitudes, such as the case shown in Fig.~\ref{fig:hx_num}(c) and (d).

Within this framework, we assume that the whole drop translates with constant speed $\tilde{U}$, so that
\begin{equation}
\tilde{x}_{\rm r}(\tilde{t}) = \tilde{U} \tilde{t}, \qquad \tilde{x}_{\rm f}(\tilde{t}) = \tilde{\rm{w}} + \tilde{U}\tilde{t},
\end{equation}
where (constant) $\tilde{\rm{w}}$ is the width of the moving drop. Here, we will consider that the rear (front) contact lines of the traveling drop are given by $\tilde{x}_1^\ast$ ($\tilde{x}_2^\ast$) (see Fig.~\ref{fig:scheme}), since the SAW force that drives it is felt only where $\tilde{h}> \tilde{h}^\ast$. Then, we define $\tilde{\xi} = \tilde{x} - \tilde{x}_1^\ast( \tilde{t} )= \tilde{x} - \tilde{U} \tilde{t}$ and assume that $\tilde{h}(\tilde{x},\tilde{t})=\tilde{h}(\tilde{\xi} )$, so that Eq.~(\ref{eq:dhdtadim}) becomes (see also \cite{buckingham_2003,perazzo_2004,giacomelli_2016} for similar studies without SAW forces)
\begin{equation}
\label{eq:dhdxi}
\tilde{U} \frac{d \tilde{h}}{d \tilde{\xi}} + \frac{d }{d \tilde{\xi}}  \left[ \tilde{h}^3 \frac{d}{d \tilde{\xi}}\left( -\frac{d^2 \tilde{h}}{d {\tilde{\xi}}^2} + {\rm Bo} \,  \tilde{h} + \frac{\mathcal{S} \tilde{C}_z}{2 \tilde{K}_z} \psi(\tilde{\xi} ,\tilde{h}) \right) 
 + \tilde{\mathcal{C}} \frac{3\mathcal{S}}{8 \tilde{K}_z^4}\psi( \tilde{\xi},\tilde{h}) \bigg(2 \tilde{K}_z^2 \tilde{h}^2-1+e^{2 \tilde{K}_z \tilde{h}}(1-2 \tilde{K}_z \tilde{h}) \bigg)\right] = 0,
\end{equation}
where
\begin{equation}
    \psi(\tilde{\xi} ,\tilde{h})= e^{-2 [ \tilde{k}_{\rm s,{\rm i}} \tilde{\xi} + \tilde{K}_z (\tilde{h}- \tilde{h}^\ast)]}.
\end{equation}
This equation can be integrated once to yield
\begin{equation}
\label{eq:dhdxi1}
\tilde{U} \tilde{h} + \tilde{h}^3 \frac{d }{d \tilde{\xi}}  \left[ -\frac{d^2 \tilde{h}}{d {\tilde{\xi}}^2} + {\rm Bo} \,\tilde{h} + \frac{\mathcal{S} \tilde{C}_z}{2\tilde{K}_z} \psi(\tilde{\xi}, \tilde{h}) \right] + \tilde{\mathcal{C}} \frac{3\mathcal{S}}{8 \tilde{K}_z^4} \psi(\tilde{\xi},\tilde{h}) \bigg(2 \tilde{K}_z^2 \tilde{h}^2-1+e^{2 \tilde{K}_z \tilde{h}}(1-2 \tilde{K}_z \tilde{h})\bigg)= \tilde{J},
\end{equation}
where $\tilde{J}$ represents the flux. The traveling wave solution must be calculated for $0 \leq \tilde{\xi} \leq \tilde{\rm w}$, along with the following boundary conditions at $\tilde{\xi} =0$ 
\begin{equation}
\tilde{h} (0)= \tilde{h}^\ast, \qquad \tilde{h}^\prime (0)= \tilde{h}^{\prime \prime \prime}(0)=0,
\end{equation} 
where the prime denotes $d/d \tilde{\xi}$. The values of $\tilde{h}^{\prime \prime}(0)$, $\tilde{U}$ and $\tilde{\rm w}$ are determined by 
\begin{equation}
    \label{eq:cond}
    \tilde{h} (\tilde{\rm w})=  \tilde{h}^\ast, \quad \tilde{J}(0)= \tilde{J} (\tilde{\rm w}), \quad \tilde{A}_{\rm d}=\int_0^{\tilde{\rm w}} \tilde{h}(\tilde{\xi})\, d\tilde{\xi}.
\end{equation}
Note that $\tilde{J}$ corresponds to the flux within a thin film of thickness $\tilde{h}^\ast$ that enters the drop at $\tilde{\rm w}$ and comes out of it at $\tilde{\xi} =0$ (with respect to the reference frame fixed at the drop). 

We developed an iterative scheme to perform the numerical integration of Eq.~(\ref{eq:dhdxi1}), under the conditions given by  Eq.~(\ref{eq:cond}). We start by guessing values of $\tilde{h}^{\prime \prime}(0)$, $\tilde{U}$ and $\tilde{\xi}_{\rm f}$ to perform the integration of Eq.~(\ref{eq:dhdxi1}) and modify them accordingly until these conditions are satisfied within a small relative error (typically, $10^{-7}$ or even smaller). After obtaining convergence for a given $\tilde{A}$, we use these converged values of $\tilde{h}^{\prime \prime}(0)$, $\tilde{U}$ and $\tilde{\xi}_{\rm f}$ as new guess values for $\tilde{A} + \Delta \tilde{A}$ (numerical continuation; $\Delta \tilde{A}$ has to be very small to reach convergence). 

\begin{figure}
\centering
\subfigure[]
{\includegraphics[width=0.43\linewidth]{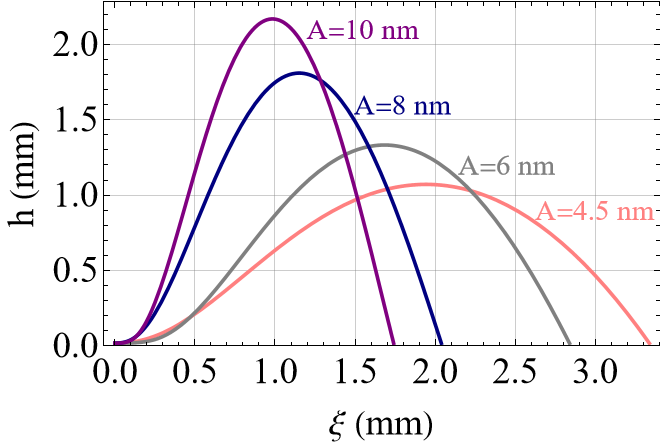}}
\hspace{0.2cm}
\subfigure[]
{\includegraphics[width=0.47\linewidth]{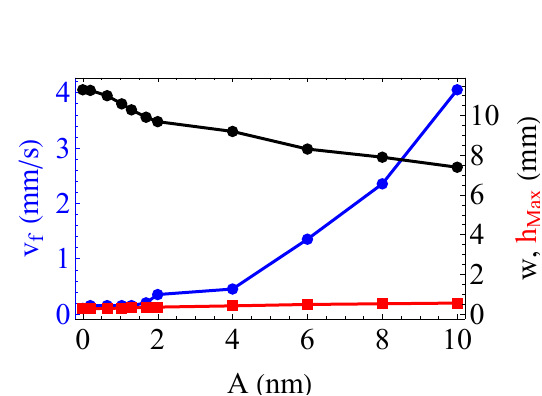}}
\caption{(a) Thickness profile of the traveling wave solution for: $A=4.5$~nm, $6$~nm, $8$~nm, $10$~nm (b) Front speed, ${\rm v}_{\rm f}$ (mm/s, {\color{blue}blue} line), drop width, $\rm w$ (mm, black line), and maximum height, $h_{\rm max}$ (mm, {\color{red}red} line), of the traveling drop as a function of $A$; front speed is plotted on the left $y$-axis while drop width and maximum height are on the right $y$-axis.}
\label{fig:drop_travel}
\end{figure}

Figure~\ref{fig:drop_travel}(a) shows the dimensional drop profiles for the values of $A$ considered so far. Clearly, drops become taller and narrower as $A$ increases. Figure~\ref{fig:drop_travel}(b) shows the dependence of drop speed, ${\rm v}_{\rm f}$, its width, $\rm w$, and the maximum thickness, $\rm h_{\rm Max}$, on $A$.
Note that this simplified model predicts that the speed depends approximately quadratically on the amplitude, $A$, while the maximum height of the drop depends approximately linearly on $A$, as was found in the full numerics (see Figure~\ref{fig:width}(b)). 

This approximate formulation shows that the simple traveling-wave solution can capture, at least qualitatively, certain features of the experimental and numerical drops. Clearly, this description cannot account for the rear region of the drop (see Figs.~\ref{fig:hx_num} and \ref{fig:drop_travel}(a)) because the values of $h'$ and $h'''$ at the left contact line cannot be determined \emph{a priori}, and both have been set to zero (for lack of better choice) in the traveling wave calculations.

\section{Conclusions}
\label{sec:conclusion}

In this work, we consider the problem of a spreading silicone oil drop driven by a surface acoustic wave (SAW) propagating in the supporting substrate. Our particular focus is on a theoretical framework that describes drops of millimetric thickness, for which the main driving mechanism is the Eckart streaming. Therefore, we focus on a different regime from that relevant for the much thinner drops considered by~\cite{Rezk12,Rezk14}, for which Schlichting and Rayleigh streaming effects are dominant. To the best of our knowledge, modeling the dynamics of millimetric drops under the action of SAW and capillary stresses has not been considered so far in the literature. We also present selected experimental results for the qualitative validation of our model.

While developing our theoretical description of drop spreading, it became apparent that an important effect, which requires careful modeling, involves the attenuation of the SAW under the evolving drop. The model we derived, assuming the thin film limit and implementing the long-wave approximation, shows that the oil drop is essentially pushed by the SAW forcing from behind.
For the highly viscous oils considered in our experiments, the effective SAW forcing is due to both viscous dissipation, and to spatial variation of the leaked acoustic wave in the fluid due to SAW attenuation in the solid.  A consequence of the driving mechanism is that the dynamics itself is substantially different from the familiar case of drops driven by gravity, where the volume force is a constant and not rapidly decreasing along the drop, as occurs with the SAW force.  In particular, we find that the drops tend to keep almost the same shape as time and spreading progress, and also that they reach essentially constant spreading speeds at long times.  This contrasts with the gravity-driven case, where the spreading speed decreases due to drop thinning. 

While not all details of the experimental results are fully understood, in general, we find reasonable agreement between the experimental and theoretical results, further supporting the basic premises underlying the developed acousto-fluidic model. In particular, our model provides good qualitative predictions for drop shapes, spreading speeds, and the dependence of the results on the SAW displacement amplitude at the solid surface (the SAW intensity). Both theoretical and experimental results show asymmetric drop shapes with a thin trailing film left behind, and both appear to reach a constant spreading speed asymptotically in time. The latter finding motivated us to formulate a simplified traveling wave model. The resulting analytical solution predicts an approximately linear increase in the drop thickness and a quadratic increase of the spreading speed with the acoustic intensity (SAW normal displacement amplitude at the solid surface), $A$. Such predicted trends are consistent with those of the full theoretical model and the experiments.  

While many features of the experimental results have been rationalized using our theoretical models and simulations, much remains to be done. Further investigation is required to shed light on the features of dynamic drop profiles under SAW excitation. The transition from three-dimensional (3D) to quasi-two-dimensional (2D) dynamics is also intriguing: under strong SAW excitation, a drop spreads primarily along the path of the SAW and may be described to leading order using the 2D model given here. However, at weak SAW excitation, where the drop also spreads radially due to the relaxation of capillary stress at the free drop surface, the drop dynamics is inherently of the 3D type. Hence, 3D aspects of drop dynamics, in particular in cases where the contributions to flow from capillary and SAW stresses are comparable, remain to be simulated and understood.  

\appendix

\section{First-Order Velocity Components and SAW Force} \label{app:RealVel}
As mentioned in the main text, although we deal with imaginary equations in the model derivation, it is understood that the real part of these complex equations is taken as the result for physically-relevant real quantities. This includes the first-order velocity field, $\mathbf{v}_1$, given by Eq.~\eqref{eq:velocity1}, and the acoustic force on the induced steady streaming flow, $\mathbf{F}_{\rm s}$, given by Eq.~\eqref{eq:Fs}, which we now discuss in detail.

We first examine $\mathbf{v}_1$ whose real parts in two dimensions can be written as
\begin{align}
    v_{1,x} &= V_xA\omega e^{-k_{\rm s,{\rm i}} x-K_zz}, \label{eq:u} \\
    v_{1,z} &= V_zA\omega e^{-k_{\rm s,{\rm i}} x-K_zz},
    \nonumber
\end{align}
where the factors $V_x$ and $V_z$ are
\begin{eqnarray}
    V_x &=& \frac{1}{\Delta} \left[ k_{\rm s,{\rm i}} \cos \left(\frac{\theta }{2}+k_{\rm s,{\rm r}} x+ z \Delta \sin \left(\frac{\theta }{2}\right)- \omega t  \right)+k_{\rm s,{\rm r}} \sin \left(\frac{\theta }{2}+k_{\rm s,{\rm r}} x+ 
    z\Delta \sin \left(\frac{\theta}{2} \right)-\omega t \right) \right],\\
    V_z &=& \cos \left(k_{\rm s,{\rm r}} x + z \Delta \sin \left(\frac{\theta }{2}\right)-\omega t \right),
\end{eqnarray}
with 
\begin{align}
    \Delta &= \sqrt{\left| \kappa_{\rm s}^2 - \kappa^2 \right|}=\sqrt[4]{\left(k_{\rm s,{\rm i}}^2-k_{\rm s,{\rm r}}^2+k_{\rm r}^2-k_{\rm i}^2\right)^2+(2 k_{\rm s,{\rm i}} k_{\rm s,{\rm r}}-2 k_{\rm i}  k_{\rm r})^2}, \label{eq:Delta}\\
    \theta &= \arg \left( \kappa_{\rm s}^2 - \kappa^2 \right),\label{eq:theta}\\
    K_z &= \Delta \cos \left( \theta / 2 \right).\label{eq:kz}
\end{align}

Turning next to $\mathbf{F}_{\rm s}$, in order to find the real part we substitute Eq.~\eqref{eq:u} with the subsequently defined parameters into Eq.~(\ref{eq:Fs}), and find explicit expressions for the acoustic force components in the form,
\begin{eqnarray}
    F_{{\rm s},x} &=& C_x P_0 \, e^{-2 (k_{\rm s,{\rm i}} x + K_z z)}, 
\label{eq:Fsx}\\
    F_{{\rm s},z} &=& C_z P_0 \, e^{-2 (k_{\rm s,{\rm i}} x + K_z z)},
\label{eq:Fsz}
\end{eqnarray}
where $P_0$ is defined in Eq.~(\ref{eq:psi}), and
\begin{eqnarray}
    C_x &=& \frac{k_{\rm s,{\rm i}} k_{\rm s,{\rm r}}^2+\Delta  \cos \left(\frac{\theta }{2}\right) k_{\rm s,{\rm i}} K_z+k_{\rm s,{\rm i}}^3+\Delta  \sin \left(\frac{\theta }{2}\right) k_{\rm s,{\rm r}} K_z}{\Delta ^2},\label{eq:Cx}\\
    C_z &=& \frac{\cos \left(\frac{\theta }{2}\right) k_{\rm s,{\rm i}}^2+\sin \left(\frac{\theta }{2}\right) k_{\rm s,{\rm i}} k_{\rm s,{\rm r}}+\Delta \, K_z}{\Delta }. \label{eq:Cz}
    \end{eqnarray}

We point out that $\mathbf{F}_{{\rm s}}$ is a non--conservative force; it cannot be written as the gradient of a potential, except for a special choice of parameter values that satisfy $k_{\rm s,{\rm i}} C_z = K_z C_x$ (see Eqs.~\eqref{eq:Fsx} and \eqref{eq:Fsz}). 

\section{Approximate solution of the SAW force}\label{app:LeakySAW} 

In this section, we discuss an approximate solution for the SAW force in certain contexts.  Section~\ref{app:oil} discusses simplifications to the model considered in Section~\ref{app:RealVel}, using the appropriate parameters for PDMS, as considered in the experiments and in the main body of the paper.  To illustrate the differences that occur in the case of less viscous fluid (water), in Sec.~\ref{app:water} we discuss appropriate simplifications for such cases. Finally, in Sec.~\ref{app:inviscid} we discuss the model that results if viscous losses are neglected, and the relation of the obtained model to that proposed decades ago by~\citet{shiokawa1990}.

\subsection{Oil (PDMS)}
\label{app:oil}
In this section, we take a closer look at the acoustic force $\mathbf{F}_{\rm s}$, given by Eq.~\eqref{eq:Fs}, on the induced steady streaming flow. We first note that, for our parameter values, we can define the small quantities
\begin{equation}
    \varepsilon_{\rm s} = \frac{k_{\rm s,{\rm i}}}{k_{\rm s,{\rm r}}} \qquad\text{and}\qquad 
    \varepsilon=\frac{k_{\rm i}}{k_{\rm r}}\approx\frac{\beta}{2}.
    \label{eq:epsis}
\end{equation}
These two parameters $\varepsilon_{\rm s}$ and $\varepsilon$ are always small, and of the same order of magnitude, for the PDMS used here (see Eq.~\eqref{eq:eta_zeta_approx} and the surrounding discussion).
To approximate the SAW force, we can expand the auxiliary variables and SAW force coefficients, given in Eqs.~(\ref{eq:Delta}), (\ref{eq:theta}), (\ref{eq:kz}), (\ref{eq:Cx}) and (\ref{eq:Cz}), in terms of $\varepsilon_{\rm s}$ and $\varepsilon$. Retaining only linear terms in the expansions (neglecting any terms
quadratic in $\varepsilon_{\rm s}, \varepsilon$
that are of higher order than those retained) yields the following expressions
\begin{eqnarray}
\Delta &\approx&  k_{\rm s,{\rm r}} \alpha_1, \nonumber\\
\theta &\approx& \pi - \frac{2}{\alpha_1^2}(1+\alpha_1^2)\varepsilon + \frac{2}{\alpha_1^2}\varepsilon_{\rm s}, \nonumber\\
K_z &\approx & \frac{k_{\rm s,{\rm r}}}{\alpha_1}(1+\alpha_1^2)\varepsilon-\frac{k_{\rm s,{\rm r}}}{\alpha_1}\varepsilon_{\rm s}, \label{eq:Dthetakzapp} \\ 
C_x &\approx &  \frac{k_{\rm s,{\rm r}}(1+\alpha_1^2)}{\alpha_1^2}\varepsilon, \nonumber \\
C_z &\approx & \alpha_1 C_x, \nonumber
\end{eqnarray}
where we have simplified the expressions by defining
\begin{equation}
\label{eq:alpha1}
\alpha_1 = \sqrt{\frac{k^2}{k_{\rm s,{\rm r}}^2}-1} = \sqrt{\frac{c_{\rm s}^2}{c^2}-1} ,
\end{equation}
since $k_{\rm s,{\rm r}} =\omega/c_{\rm s}$.  This coefficient yields the Rayleigh angle $\theta_R$ as: $\tan \theta_R = 1/\alpha_1$.

From the outlined approximation, we obtain that the resultant force is given by
\begin{align} 
    F_{{\rm s},x} &\approx \frac{\rho_0 A^2 \omega^2 (1+\alpha_1)^2 k_{\rm s,{\rm r}}}{\alpha_1^2}\varepsilon \ e^{-2(k_{\rm s,{\rm i}}x+K_z z)}, \label{eq:fullApproxX} \\
    F_{{\rm s},z} &\approx \alpha_1 F_{{\rm s},x} .\label{eq:fullApproxZ}
\end{align}
The approximations made in Eqs.~\eqref{eq:Dthetakzapp} lead to an attenuation factor $K_z$ that is approximately $12$\% smaller and force coefficients $C_x$ and $C_z$ that are approximately $2$\% bigger and $6$\% smaller, respectively, with respect to the quantities in the exact expressions given in Eqs.~\eqref{eq:kz}, ~\eqref{eq:Cx}, and~\eqref{eq:Cz}. Note that the approximate force components obtained in Eqs.~(\ref{eq:fullApproxX}), (\ref{eq:fullApproxZ}) (also) do not yield a conservative force, since $K_z \neq \alpha_1 k_{\rm s,{\rm i}}$.

\subsection{Water}
\label{app:water}
Next, we consider modeling a lower viscosity fluid such as water, characterized by a similar density and phase velocity to oil (PDMS) so that the imaginary component of the wavenumber in the solid is similar (see Eq.~\eqref{eq:ki}). However, the large decrease in viscosity in the case of water leads to a much reduced value of the imaginary wavenumber in the liquid, $k_{\rm i}$ (see Eq.~\eqref{eq:kappa-def}). Thus for water, our assumption that $\varepsilon$ and $\varepsilon_{\rm s}$ are of the same order of magnitude fails; in fact, here we have $\varepsilon_{\rm s}^2 \sim \varepsilon$ so that we need to also retain terms that are quadratic in $\varepsilon_{\rm s}$; the next largest terms in the expansion are those proportional to $\varepsilon_{\rm s}^3$ which we also include for the purpose of comparison to the approximation in the case of neglecting viscous dissipation which is discussed later in Sec.~\ref{app:inviscid}. Note that we neglect any cross terms that may appear in the expansions as they are higher order than the terms retained.

Performing an asymptotic expansion of the auxiliary variables and SAW force coefficients appropriate for the water leads to the following approximate expressions
\begin{align}
    \Delta &\approx k_{\rm s,{\rm r}}\alpha_1 + \varepsilon_{\rm s}^2 \bigg(\frac{k_{\rm s,{\rm r}}}{\alpha_1^3}+\frac{k_{\rm s,{\rm r}}}{2\alpha_1} \bigg), \nonumber \\
    \theta &\approx - \pi + \frac{2}{\alpha_1^2}\varepsilon_{\rm s} - \frac{2(1+\alpha_1^2)}{\alpha_1^2}\varepsilon, \nonumber \\
    K_z &\approx \frac{k_{\rm s,{\rm r}}}{\alpha_1}\varepsilon_{\rm s}-\frac{k_{\rm s,{\rm r}}(1+\alpha_1^2)}{\alpha_1}\varepsilon, \label{eq:waterApprox} \\
    C_x &\approx \frac{k_{\rm s,{\rm r}}(1+\alpha_1^2)}{\alpha_1^2}\bigg(\varepsilon+\frac{\varepsilon_{\rm s}^3}{\alpha_1^2} \bigg) , \nonumber \\
    C_z &\approx \frac{k_{\rm s,{\rm r}}(1+\alpha_1^2)} {\alpha_1}\bigg(-\varepsilon+\frac{\varepsilon_{\rm s}^3}{\alpha_1^4} \bigg). \nonumber
\end{align}
We pause to discuss the difference between the approximated complex arguments in the case of oil and water.  Note that the two expressions for $\theta$ (Eqs.~\eqref{eq:Dthetakzapp} and~\eqref{eq:waterApprox}) have the same first-order correction terms with respect to the small parameters $\varepsilon$ and $\varepsilon_{\rm s}$; however, the sign of the sum of these two terms is opposite for the case of PDMS (negative) and water (positive) - which is due to the much reduced value of the imaginary wavenumber in the liquid, $k_{\rm i}$, in the case of water as compared to PDMS (see Eqs.~\eqref{eq:kappa} and~\eqref{eq:tan_beta}). The result of this sign change has the effect of moving the complex argument from quadrant II (in the case of PDMS) to quadrant III (in the case of water); as a result, the approximation for the sine function in the SAW force coefficients changes sign and thus the overall value of the SAW force coefficients is much reduced (see Eqs.~\eqref{eq:Cx} and~\eqref{eq:Cz}).

From the approximation in Eq.~\eqref{eq:waterApprox}, we can see that the resultant force for water, at leading order, is given by:
\begin{align} 
    F_{{\rm s},x} &\approx \frac{\rho_0 A^2 \omega^2 (1+\alpha_1)^2 k_{\rm s,{\rm r}}}{\alpha_1^2}\varepsilon \ e^{-2(k_{\rm s,{\rm i}}x+K_z z)}, \label{eq:waterApproxFx} \\
    F_{{\rm s},z} &\approx -\alpha_1 F_{{\rm s},x} \label{eq:waterApproxFz}.
\end{align}
Using appropriate parameter values for water, we find that the SAW force for water, Eqs.~\eqref{eq:waterApproxFx} and~\eqref{eq:waterApproxFz}, has a coefficient that is approximately two orders of magnitude smaller than that for PDMS, see Eqs.~\eqref{eq:fullApproxX} and~\eqref{eq:fullApproxZ} (due to the fact that $\varepsilon$ is much smaller for water than oil). Additionally, the $z$ component of the SAW force has a negative prefactor, $C_z$. 

Before closing this discussion, we note that if one wants to accurately study the dynamics of water drops in the present context, one must also include a disjoining pressure term in Eq.~\eqref{eq:P2} to account for the partial wettability of the fluid. 

\subsection{Neglecting Viscous Dissipation}
\label{app:inviscid}
We now discuss briefly a modified model obtained if viscous dissipation is neglected by setting $k_{\rm i}=0$, which is akin to the approach presented by~\citet{shiokawa1990,Shiokawa:1989tg}. To do this, the simplest approach is to set $k_{\rm i}=0$ in the exact formulae (Eqs.~\eqref{eq:Delta},~\eqref{eq:theta},~\eqref{eq:kz},~\eqref{eq:Cx}, and~\eqref{eq:Cz}) and then re-expand the quantities in terms of the remaining small parameter $\varepsilon$.
Doing this, one finds that the auxiliary variables and the SAW force coefficients are now given by
\begin{eqnarray}
\Delta &\approx&  k_{\rm s,{\rm r}} \alpha_1, \nonumber\\
\theta &\approx& -\pi + \frac{2}{\alpha_1^2}\varepsilon_{\rm s}, \nonumber\\
K_z &\approx & \frac{k_{\rm s,{\rm r}}}{\alpha_1}\varepsilon_{\rm s} \label{eq:Dthetakzapp1}, \\
C_x &\approx &  \frac{k_{\rm s,{\rm r}}(1+\alpha_1^2)}{\alpha_1^4}\varepsilon_{\rm s}^3,\nonumber \\
C_z &\approx & \frac{1}{\alpha_1} C_x. \nonumber
\end{eqnarray}
We note that this approximation is exactly what is found by simply setting $\varepsilon=0$ in the approximation for the case of a low viscosity fluid such as water (see Eq.~\eqref{eq:waterApprox}).
Hence, if viscous dissipation is neglected, the SAW force is given as
\begin{align}
    F _{{\rm s},x} &\approx \frac{\rho A^2 \omega^2 (1+\alpha_1^2)}{\alpha_1^4} k_{\rm s,{\rm r}} \varepsilon_{\rm s}^3 e^{-2k_{\rm s,{\rm i}}(x+\frac{1}{\alpha_1} z)}, \label{eq:fullApproxNokiX} \\
    F_{{\rm s},z} &\approx\frac{1}{\alpha_1} F_{{\rm s},x} ,\label{eq:fullApproxNokiZ}
\end{align}
which can be rewritten in terms of a potential $\phi_{\rm s}$, so that $\vec{F}_{\rm s}=-\nabla \phi_{\rm s}$ is a conservative volume force with
\begin{equation}
    \phi_{\rm s} = \frac{\rho A^2 \omega^2 (1+\alpha_1^2)}{2\alpha_1^4 }\varepsilon_{\rm s}^2 e^{-2k_{\rm s,{\rm i}}(x+\frac{1}{\alpha_1}z)}~.
    \label{eq:phis}
\end{equation}
We note that without viscous dissipation, the force components are five orders of magnitude smaller than when the viscosity is considered, compare for example Eqs.~\eqref{eq:Cx},~\eqref{eq:Cz} to Eq.~\eqref{eq:Dthetakzapp1} with the given parameter values. As a result, the force in Eqs.~\eqref{eq:fullApproxNokiX} and~\eqref{eq:fullApproxNokiZ} has no effect on drop dynamics on the timescales investigated.  Such a finding inspires comparison with the results  presented in~\citet{shiokawa1990,Shiokawa:1989tg}, and given in terms of our variables as
\begin{eqnarray}
    \phi_{\rm shio} = \frac{1+\alpha_1^2}{2}\rho A^2 \omega^2 e^{-2k_{\rm s,{\rm i}}(x+\alpha_1 z)}.
\end{eqnarray}
Note that, for the considered parameter values, this expression leads to a force that is approximately three orders of magnitude {\em larger} compared to what is found from our model, i.e., Eq.~\eqref{eq:phis}. Further, our model predicts a different attenuation factor in $z$ as well as a different force prefactor than what is found in~\citet{shiokawa1990,Shiokawa:1989tg}. In those works, the authors find that $K_z=\alpha_1 k_{\rm s,{\rm i}}$, as opposed to our model which yields $K_z\approx {k_{\rm s,{\rm r}}\varepsilon_{\rm s}/\alpha_1}=k_{\rm s,{\rm i}}/\alpha_1$. From the physical point of view, one expects larger viscous dissipation at a given $z$ (distance from the substrate, into the liquid), when the Rayleigh angle, $\theta_R$, increases and tends to $\pi/2$ (see Fig.~\ref{fig:scheme}, where the angle is measured clockwise from the normal to the substrate). This is because the leaky wave travels a longer distance along the ray (for given $z$) for larger $\theta_R$, and thus its amplitude is more attenuated. For example, if $c \rightarrow c_{\rm s}$, we have $\theta_R \rightarrow \pi/2$ and $\alpha_1=1/\tan \theta_R \rightarrow 0$ (from Snell's law, we have $\sin(\theta_R)=c/c_{\rm s} <1$). Since $K_z$ must increase for $\theta_R \rightarrow \pi/2$, we have $\tan \theta_R \rightarrow \infty$, and then one expects $K_z \propto 1/\alpha_1 \rightarrow  \infty$, and not $K_z \propto \alpha_1 \rightarrow 0$ as in \cite{shiokawa1990,Shiokawa:1989tg}. 

\section{Computational implementation}
\label{app:comsol}
Equation~(\ref{eq:dhdtadim}) is written in a form convenient for the use of COMSOL\texttrademark    Multiphysics PDE Coefficients Form. This package solves, by finite elements, a vectorial equation for the unknown vector $\vec u=(u_1,u_2,\dots,u_N)^\top$.  The equation is of the form
\begin{equation}
\label{eq:gral}
{\bf e} \frac{\partial^2 \vec u}{\partial t^2}+{\bf d}  \frac{\partial \vec u}{\partial t} + \nabla \cdot 
\left( - {\bf c}  \nabla \vec u - {\bf \alpha} \,  \vec u + {\bf \gamma}\right) + {\bf \beta}  \, \nabla \vec u + {\bf a} \, \vec u = \vec f,
\end{equation}
where the coefficients of the $N$ scalar equations are in the matrices ${\bf e}$, ${\bf d}$, ${\bf \gamma}$, ${\bf a}$ (of dimensions $N\times N$), ${\bf \alpha}$, ${\bf \beta}$ (of dimensions $N\times N\times n$), ${\bf c}$ (of dimensions $N\times N\times n \times n$) and the vector $\vec f$ (of dimension $N$), where $n$ is the spatial dimension of the problem ($n=1,2,3$). In index notation, this equation reads as
\begin{equation}
e_{ij}\frac{\partial^2 u_j}{\partial \tilde{t}^2}+d_{ij}\frac{\partial u_j}{\partial \tilde{t}}+\frac{\partial}{\partial \tilde{x}_l} \left( -c_{ijkl} \frac{\partial u_j}{\partial \tilde{x}_k}
-\alpha_{ijl} u_j + \gamma_{il} \right)+ \beta_{ijl} \frac{\partial u_j}{\partial \tilde{x}_l} +
a_{ij} u_j = f_i,
\end{equation}
where $i,j=1.\ldots,N$ and $k,l=1,\ldots,n$.

The considered system, given by Eq.~(\ref{eq:dhdtadim}), contains two components $u=(\tilde{h}, \tilde{\cal P})$ ($N=2$) and we use two equations, namely, Eqs.~(\ref{eq:dhdtadim}) and (\ref{eq:Pad}) for ($n=1)$, corresponding to the solution depending on a single spatial variable, $\tilde{x}$. 

In the following, we list the non-vanishing coefficients (we omit the indexes $k$ and $l$ for brevity and consider $\tilde{x}_1\equiv \tilde{x}$ since $k=l=1$): 
\begin{itemize}
\item Row $1$ ($i=1$) for Eq.~(\ref{eq:dhdtadim})
\begin{equation}
d_{11}=1,\quad c_{12}= \tilde{h}^3, \quad 
\gamma_1 =- \tilde{\cal C} \frac{3 {\cal S}}{8 \tilde{K}_z^4} \psi(\tilde{x}, \tilde{h}) \Big(2 \tilde{K}_z^2 \tilde{h}^2-1+ e^{2 \tilde{K}_z \tilde{h}}(1-2 \tilde{K}_z \tilde{h}) \Big) \bigg]
\end{equation}
\item Row $2$ ($i=2$) for Eq.~(\ref{eq:Pad})
\begin{equation}
c_{21} = -1, \quad a_{21}=-{\rm Bo},\qquad a_{22}=1, \qquad 
f_2=\frac{{\cal S} \tilde{C}_z}{2 \tilde{K}_z} \psi(\tilde{x}, \tilde{h})
\end{equation}
\end{itemize}
At the domain ends, we apply Dirichlet boundary conditions $\tilde{h}= \tilde{h}_{\rm p}$, where $\tilde{h}_{\rm p}$ is the precursor film thickness, and $\partial \tilde{h}/\partial \tilde{x}=0$. In order to achieve convergent numerical simulations, we discretize the domain using $\Delta \tilde{x} =\tilde{h}_{\rm p}$.

\vspace{0.2in}
{\bf Acknowledgments}

This work was supported by the donors of ACS Petroleum Research Fund under PRF\# 62062-ND9 and by BSF grant No. 2020174. J.A.D acknowledges support from Consejo Nacional de Investigaciones Científicas y Técnicas (CONICET, Argentina) with Grant PIP 02114-CO/2021 and Agencia Nacional de Promoción Científica y Tecnológica (ANPCyT, Argentina) with Grant PICT 02119/2020.

\vspace{0.2in}
{\bf Declaration of Interests}

The authors report no conflict of interest. 

\bibliographystyle{jfm}
\bibliography{films.bib}

\end{document}